\documentclass[12pt,preprint]{aastex}
\usepackage{natbib,graphicx}

\shorttitle{Acceleration History of the Universe}
\shortauthors{Daly et al.}

\begin{document}

\title{Improved Constraints on the Acceleration History of the Universe and the Properties of the Dark Energy}
\author{Ruth A. Daly\altaffilmark{~}}
\affil{Department of Physics, Penn State University, Berks Campus, P. O. 
Box 7009, Reading, PA 19610}
\email{rdaly@psu.edu}
\author{S. G. Djorgovski\altaffilmark{~}}
\affil{Division of Physics, Mathematics, and Astronomy, 
California Institute of Technology, MS 105-24, 
Pasadena, CA 91125}
\author{Kenneth A. Freeman, Matthew P. Mory}
\affil{Department of Physics, Penn State University, Berks Campus, P. O. 
Box 7009, Reading, PA 19610}
\author{C. P. O'Dea, P. Kharb, \& S. Baum\altaffilmark{~}}
\affil{Rochester Institute of Technology, 54 Lomb Memorial Drive, 
Rochester, NY 14623}


\eject

\begin{abstract}
We extend and apply a model-independent analysis method developed earlier by 
Daly \& Djorgovski to new samples of supernova standard candles, 
radio galaxy and cluster standard rulers, and use it to constrain physical
properties of the dark energy as functions of redshift.
Similar results are obtained for the radio galaxy and supernova 
data sets, which rely upon completely independent methods, 
suggesting that systematic errors are relatively small for
both types of distances; distances to SZ clusters show a scatter 
which cannot be explained by the quoted measurement errors. 
The first and second derivatives of the distance are 
compared directly with predictions in a standard model
based on General Relativity.  The good agreement indicates
that General Relativity provides an accurate description of
the data on look-back time scales of about 
ten billion years. 
The first and second derivatives are combined to
obtain the acceleration parameter $q(z)$, assuming only 
the validity of the Robertson-Walker metric, 
independent of a theory of gravity and of the physical nature of the dark energy.  
The data are analyzed using a sliding window fit and 
using fits in independent redshift bins.  
The acceleration of the universe at the
current epoch is indicated by the sliding window fit
analysis.  
The effect of non-zero space
curvature on $q(z)$ is explored; for a plausible range of values of $\Omega_k$
the effect is small and 
causes a to shift to the redshift at which the universe
transitions from deceleration to acceleration.
We solve for the pressure, energy density, equation
of state, and potential and kinetic energy of the dark energy
as functions of redshift assuming that General Relativity is
the correct theory of gravity.  Results obtained
using a sliding window fit indicate that a cosmological constant in 
a spatially flat universe provides a good description of
each of these quantities over the redshift range from zero to about
one. We define a new function, the dark energy indicator, in terms
of the first and second derivatives of the coordinate distance  
and show how this can be used
to measure deviations of $w$ from $-1$ and to
obtain a new and independent measure of $\Omega_m$.

\end{abstract}


\eject

\section{INTRODUCTION}

Understanding of the physical nature of the dark energy which appears to be
driving the accelerated expansion of the universe is 
among the most pressing and important topics 
in cosmology today. Studies of the 
expansion history of the universe allow us to 
constrain the physical nature of its matter and energy constituents.
One way that the expansion and acceleration history of the
universe can be studied is through the use of a set of
coordinate distances and redshifts for some standard set of objects.  
Type Ia supernovae provide a modified standard candle 
(e.g. Phillips 1993, Hamuy et al. 1995) 
that allow the
distance modulus, luminosity distance, and 
coordinate distance to each source to be determined.
The recent data sets presented by Astier et al. (2006),
Riess et al. (2007), Wood-Vasey et al. (2007), and
Davis et al. (2007) have been analyzed by these groups
and compared with numerous models by other researchers.  

In a novel, largely model-independent approach to this problem, it
was shown by Daly \& Djorgovski (2003) that the first and
second derivatives of the coordinate distance with respect
to redshift could be obtained from the coordinate distances
and combined to solve for the expansion rate $H(z)/H_0$ and
acceleration rate $q(z)$ of the universe. The 
functions $H(y^{\prime})$ and $q(y^{\prime},y^{\prime \prime})$
are exact, that is, they are not obtained by expansions in 
terms of derivatives about some point. 
The only assumptions are that the universe is described by 
a Robertson-Walker metric and has zero space curvature. 
The results are independent of the contents of the 
universe and their physical properties,
and even independent of whether General Relativity 
provides an accurate description of the universe. 
Here, we drop the assumption of 
zero space curvature; it turns out that the deceleration parameter
at a redshift of zero, $q_0$, remains the same, independent
of whether space curvature is zero or not. 

In this paper we expand on the previous analysis done by 
Daly \& Djorgovski (2003, 2004).  First, we use updated and expanded 
data sets, as described in section 2.1. Second, we introduce
a more direct way to compare the model-independent results 
obtained from the data with predictions; this is done 
by directly comparing
the first and second derivatives of the coordinate distance
with respect to redshift 
to predicted values in various models, as described in 
section 2.2. Third, we analyze the data using both a
sliding window fit and fits in independent redshift bins.  
To solve for the physical properties of the dark energy as functions
of redshift, a theory of gravity must be specified.  
To determine the properties of the dark energy, General
Relativity is taken to be the correct theory of gravity, allowing
us to solve for the pressure, energy density, and equation of 
state of the 
dark energy as a function of 
redshift  in section 2.4.  Fourth, in section 2.4, we introduce a way 
to solve for the potential and kinetic energy densities of the dark energy
as functions of redshift. In addition, we define a new function,
the dark energy indicator, which provides a measure of deviations
of $w$ from $-1$ and a new and independent measure of $\Omega_m$.
Fifth, in section 2.3, these derivatives are combined
to solve for the expansion and acceleration rates of the 
universe as functions of redshift for both zero and non-zero
space curvature; in our previous work we have not considered
the effects of non-zero space curvature. 
The only assumption that must be made to 
obtain the functions $H(z)/H_0$ and $q(z)$ from the data are 
that the Robertson-Walker metric is valid in our universe. 
A discussion and conclusions follow 
in section 3.

\section{DATA AND ANALYSIS}

\subsection{Data sets used}

We consider three types of distances: those determined from luminosity
distances to supernova standard candles (SN), those determined from
the angular diameter distances to radio galaxies (RG), and those
determined to clusters of galaxies with SZ measurements of angular
diameter distances (CL).  

The SN samples include those of Davis et al. (2007),
Riess et al. (2007), and Astier et al. (2006); these authors provide 
the pertinent details about their measurements.
There is some overlap between these supernovae samples, and
this comparison allows the effects of different samples 
and sub-samples to be seen. 
In addition,
a comparison between the values of $y(z)$, $y^{\prime}(z)$, 
and $y^{\prime \prime}(z)$ for the different samples goes hand in
hand with a comparison of the best fit parameter values obtained
in different models for these same samples, 
described by Daly et al. (2007).  
The 71 new supernovae presented by Astier et al. (2006) are
included in both the Riess et al. (2007) and Davis et al. (2007) samples,
and the high-redshift HST supernovae of Riess et al. (2007) are 
included in Davis et al. (2007) sample which, otherwise, includes
only ESSENCE supernovae and low-redshift supernovae. These 
comparisons can be quite helpful, as illustrated by the 
work on Nesseris \& Perivolaropoulos (2006). 

The dimensionless coordinate distances $y$ to supernovae
can be  
obtained from the published distance moduli $\mu$ using the best
fit value of $\kappa_{SN}$ and the relations 
$\mu = \kappa_{SN}+5log10[y(1+z)]$
and $\sigma_y =y \sigma_{\mu} (ln(10)/5)$. There are several
ways to determine $\kappa_{SN}=25-5log10(H_0/c)$ for 
published data sets that do not indicate the effective value of 
$H_0$ adopted to obtain $\mu$; the different 
methods provide values of $\kappa_{SN}$ that are in good
agreement.   
Here, we use the best fit value
of $\kappa_{SN}$ obtained by Daly et al. (2007).  
In the cases where a value of $H_0$ is included in
the publication of the values of $\mu$ (e.g. Astier et al.
2006) the value of $H_0$ adopted by Astier et al. (2006)
is recovered to very high accuracy.   

We use the new RG sample of Daly et al. (2007); eleven new 
radio galaxies were observed and analyzed, which increases the 
sample size to 30 radio galaxies with redshifts between zero
and about 1.8.  

Finally, we also use the angular diameter distances to a sample of
38 clusters determined with the SZ measurements by 
Bonamente et al. (2006). The angular diameter distances $d_A$
obtained by Bonamente et al. (2006) for the hydrostatic equilibrium
model were used.  To convert from the angular
diameter distance to the dimensionless coordinate distance
we  need to remove the value of $H_0$ that was adopted
by Bonamente et al. (2006), so we use 
their best fit value of $H_0$ of 
$76.9 {}^{+3.9}_{-3.4}$ to obtain the dimensionless coordinate
distance $y$ to each of their clusters using the well-known relations
$d_A = (a_0r)/(1+z)$ and $y = (H_0/c)(a_0r)$.  

After a detailed comparison between the SN and RG samples, we
combine the Davis et al. (2007) supernovae sample with the
Daly et al. (2007) radio galaxy sample, and study the combined
sample of 222 sources. 
Dimensionless
coordinate distances $y$ and their uncertainties $\sigma(y)$
are obtained for these samples as described by Daly et al. (2007),
and are listed here in Table 1. 
We also study results obtained by adding the cluster
sample of Bonamente et al. (2006) to obtain a sample of 260
sources, and these distances are also included in Table 1.   
Figure 1 shows a comparison of the distances for these three data sets
relative to the standard Lambda Cold Dark Matter (LCDM) model
with $\Omega_m=0.3$ and $\Omega_{\Lambda}=0.7$.  
A comparison of each data set with the standard LCDM model 
indicates 
that both the SN and RG data sets provide reliable cosmology probes,
while the SZ Cluster method perhaps need some refinement; the reduced
chi-square for the SN and RG is about one, as expected from the quoted
measurement errors, whereas that for the 
SZ Clusters is greater than two, suggesting that the quoted errors 
substantially underestimate the true uncertainties of these 
distance measurements. However, to illustrate how our
method can be applied to coordinate distances obtained using
different methods, we consider the analysis of the full sample of
260 sources as well as the analysis of the sample of 222 SN and RG.  

\subsection{Determinations of $y^{\prime}$ and $y^{\prime \prime}$}

The distances $y$ to each source 
are used to obtain the distance $y(z)$ to any redshift within the 
redshift range of the sample, and first
and second derivatives of the distance with respect to redshift,
$y^{\prime}(z)$ and $y^{\prime \prime}(z)$ and their uncertainties
using the method of Daly \& Djorgovski (2003, 2004).  

In previous work, we have used $y^{\prime}$ and $y^{\prime \prime}$
to obtain $E(z) = H(z)/H_0$ and $q(z)$.  We then compared our
empirically determined functions $E(z)$ and $q(z)$ with predictions
in different models.  However, it is also possible to compare our
empirically determined functions 
$y^{\prime}(z)$ and $y^{\prime \prime}(z)$ 
directly with model predictions for these quantities.  
The predicted values of these quantities are labeled 
$y_p^{\prime}$ and $y_p^{\prime \prime}$.

For a universe with non-relativistic matter
with mean mass energy density 
$\Omega_m(1+z)^3$, mean dark energy density $\Omega_{DE}f(z)$, and
space curvature $k$  
in which Einstein's Equations apply, we have, in full generality, 
the predicted values of $y^{\prime}$,
$y_p^{\prime}$, 
and $y^{\prime \prime}$, $y_p^{\prime \prime}$ are given by 
\begin{equation}y_p^{\prime} = \left({1+\Omega_k~y_p^2 \over \Omega_m(1+z)^3 +\Omega_{DE}f(z)+
\Omega_k(1+z)^2}\right)^{1/2} 
\end{equation}
and
\begin{equation}y_p^{\prime \prime} = {y_p^{\prime} \over (1+z)} \left({\Omega_k~y_p~y_p^{\prime}~(1+z) \over (1+\Omega_ky_p^2)}-1.5{[\Omega_m(1+z)^3+\Omega_{DE}(1+w)f(z)+(2/3)\Omega_k(1+z)^2] \over [\Omega_m(1+z)^3+\Omega_{DE}f(z)+\Omega_k(1+z)^2]}
\right) 
\end{equation}
where $\Omega_k = -k/(H_0a_0)^2$, $\Omega_m=\rho_{0m}/\rho_{0c}$ 
is the zero redshift value of the mean mass density of non-relativistic
matter relative to the critical density,
$\Omega_m+\Omega_{DE}+\Omega_k =1$, and  
$y_p = \int_0^z y_p^{\prime} dz$ is obtained by numerically
integrating eq. (1). This derivation does not assume
that $w=P_{DE}/\rho_{DE}$, is constant, and allows for variable $w(z)$.  
The function $f(z)$ describes the
redshift evolution of the energy density of the dark energy;
in a 
quintessence  model with constant equation of state $w = P_{DE}/\rho_{DE}$, 
$f(z) = (1+z)^{3(1+w)}$,
and for a cosmological constant  $f(z) = 1 $. 

The data were analyzed using a sliding window fit, described
in section 2.2.1, and using fits in independent redshift bins, described 
in section 2.2.2.

\subsubsection{Results Obtained with a Sliding Window Fit}

Fits are done using the window $\Delta z = 0.6$ throughout
for the SN data, $\Delta z = 0.8$ for the RG data, when considered
individually, and $\Delta z = 0.6$ for the joint samples using 
the method described by Daly \& Djorgovski (2003, 2004). 
The width of the fitting window is driven by the need to obtain
useful confidence intervals for the fits by including a sufficient number
of data points.  As the size of the available data sets increases in
the future, this width could be correspondingly narrowed.
In decreasing the window function width from 0.6 to 0.5 and 0.4,
the trends and overall results remain the same, the uncertainties
increase (because there are fewer data points in the window),
and the trends become more noisy (due to sparser sampling).  
In addition, to test whether the window function has an effect
on the trends extracted from the data we created a mock data set
with the same number and redshift distribution of points as in each 
data set and the same fractional error per point, but with $y$ values
obtained from a standard LCDM cosmology with $\Omega_m = 0.3$ and
$\Omega_{\Lambda}=0.7$ and ran it through the programs to extract
$y(z)$, $y^{\prime}(z)$, and $y^{\prime \prime}(z)$.  As expected,
the uncertainties increase as the window width decreases due to the
smaller number of data points in the window.  For a window widths 
of 0.3 and 0.4 in redshift, $y(z)$, $y^{\prime}(z)$,  
$y^{\prime \prime}(z)$, and $q(z)$ match the input cosmology to 
very high precision.  For widths of 0.5 and 0.6, there is a very slight
offset of $y^{\prime \prime}(z)$ and $q(z)$ from the input cosmology 
that sets in above redshifts of about one, at
a level that is very small compared with the uncertainties.
Thus, the values of $y(z)$, $y^{\prime}(z)$, $y^{\prime \prime}(z)$, 
and $q(z)$, and quantities obtained by combining these quantities,
provide reliable determinations to redshifts well above one.  

Our completely model-independent determinations of $y(z)$,
$y^{\prime}(z)$, and $y^{\prime \prime}(z)$ are shown
in Figures \ref{yofz} through \ref{d2ydz2192}. In Figures
\ref{yofz}, \ref{dydz}, and \ref{d2ydz2}, they are compared
with the predicted value in a spatially flat universe, described by
General Relativity with
a cosmological constant $\Omega_{\Lambda}=0.7$, and non-relativistic
matter $\Omega_m = 0.3$.  This provides a reasonable description
of the data to redshift of about one or so for the supernovae
and 1.5 or so for the radio galaxies. (We note that this is not a
model fit, but simply an illustration of its compatibility with the data.)

The values of $y^{\prime}(z)$ and $y^{\prime \prime}(z)$ 
obtained for the Davis et al. (2007) supernova sample are compared with 
the best fit model parameters obtained in a 
spatially flat quintessence model, a lambda model that allows for
non-zero space curvature, and the standard LCDM model in Figures
\ref{dydz192} and \ref{d2ydz2192} 
using the best fit model parameters listed by Daly et al. (2007). 
The best fit model parameters are obtained by fitting the data 
to a model, and each model is based upon General Relativity. 
A comparison of the first and second derivatives of $y$ with 
those predicted in each of the models provides another way
to see whether the model predictions provide a good description of 
the data. The models and best fit model parameters are shown for
a lambda model that allows for non-zero space curvature, and 
a quintessence model.

Fig. \ref{d2ydz2} shows $y^{\prime \prime}$ for each of the samples.
The canonical LCDM model provides a good description of the radio
galaxy data over the redshift range from zero to 1.5.

The comparison between the values of $y^{\prime}$ and $y^{\prime \prime}$ 
determined directly from the data with those predicted in a 
standard LCDM model that relies upon the equations of General 
Relativity provides effectively a large-scale test of General Relativity. 
The good agreement obtained indicates that General Relativity
provides a good description of the data on time scales (and the corresponding
length scales) of about ten billion years. 

For the Davis et al. (2007) sample of 192 supernovae, the best fit
models track $y^{\prime}$ and $y^{\prime \prime}$ to a redshift of
about 1.2 or so, beyond which the data drops away from the model
prediction (see Figs. \ref{dydz192} and \ref{d2ydz2192}).  
Curves expected for the   
best fit parameters obtained in a quintessence model
 and
cosmological constant model with space curvature
by fitting $y(z)$, 
obtained by Daly et al. (2007), are shown as well as 
a standard LCDM model.  For $y^{\prime}$ all three curves
fit well, and some differences emerge for $y^{\prime \prime}$;
in particular, the curve with non-zero space curvature does
not fit the data well at low redshift.

\subsubsection{Results Obtained With Fits in Independent Redshift Bins}

Our sliding window fit method produces fit values which are strongly
correlated over the redshift range corresponding to the fitting window,
and are thus indicative of trends, but cannot be used simply to evaluate
a statistical significance of any particular model.  For that, we would need 
independent values at different redshifts.  To this effect, we divided the
data sample in a number of independent redshift bins (note: the data are
$not$ binned or averaged, the sample is divided in groups of points
belonging to non-overlaping redshift bins).  One drawback of this approach
is that the numbers of data points in each bin are smaller, and thus the
fit values are noisier.  Another drawback is that the boundary values of
the fits are not constrained, allowing for discontinuities in the values of
$y(z)$, $y^\prime(z)$, and $y^{\prime\prime}(z)$ at the bin boundaries, which
physically makes no sense.  This is the price of the statistical independence.

The dimensionless coordinate distances to the 
192 supernovae of Davis et al. (2007) were combined with those
to the 30 radio galaxies of Daly et al. (2007) to obtain a sample
of 222 sources with redshift between 0.016 and 1.79.  
These data were divided into bins based on the redshifts of the points 
to be able to obtain independent
determinations of $y^{\prime}$ and $y^{\prime \prime}$ and their uncertainties.
The data were divided into redshift bins with roughly equal numbers of points
per bin.  We considered 2 bins with 111 points each; 
3 bins with 74 points each; 4 bins with 55 points in the first three 
bins and 57 points in the highest redshift bin; 
and 6 bins with 37 points each. The bin, number of points per bin,
median redshift of the points in the bin, and the minimum and maximum
redshift of points within the bin are listed in Table \ref{binresults}. 
Points in each bin were 
used to determine the values of $y^{\prime}$ and $y^{\prime \prime}$ 
and their uncertainties at the median redshift of the bin. 

The results for $y^{\prime}$ are shown in Figure \ref{yprimebin} 
for 3, 4, and 6 bins; the redshift range of the points that went into
the determination of $y^{\prime}$ at the median redshift $z_{med}$ are
indicated by the horizontal lines. Three theoretical curves are included
on the figure.  The dotted line is the curve predicted by the standard
LCDM model with $\Omega_m = 0.3$, and the other two curves, which are
nearly identical are those predicted in the Cardassian model
(Freese \& Lewis 2002) and the generalized Chaplygin gas model
(Kamenshchik, Moschella, \& Pasquier 2001; Bilic, Tupper, \& Viollier 2002;
and Bento, Bertolami, \& Sen 2002) are shown for the best fit parameters
obtained by Bento et al. (2005) assuming a spatially flat universe. 

The results obtained 
for $y^{\prime \prime}$ 
are shown in Figure \ref{yprimeprimebin} for 2 and 3 bins.
The full set of results are listed in Table \ref{binresults}. 
Given the noise inherent in the data that is presently available, 
it is not possible to obtain
meaningful results for $y^{\prime \prime}$ 
with a larger number of independent bins.  With more data,
we will be able to obtain this quantity in a larger number of
independent redshift bins. 

\subsection{Determinations of $H(z)/H_0$ and $q(z)$ for Zero Space Curvature}

The dimensionless expansion rate $E(z) \equiv 
H(z)/H_0$ and the deceleration parameter
$q(z)$ can be constructed directly from the fist and second derivatives 
of the coordinate distance $y^{\prime}$ and $y^{\prime \prime}$,
as discussed in detail by Daly \& Djorgovski (2003).  
The relationship between $E(z)$ and $y^{\prime}$, and that  
between $q(z)$ and $y^{\prime}$ and $y^{\prime \prime}$ are
exact; they do not represent expansions about some point. 
The only 
requirement to derive these exact relationships
is that the most the Robertson-Walker line
element applies in our universe.  With this assumption alone,
it can be shown that 
\begin{equation}
H(z)/H_0 = (y^{\prime})^{-1}~(1+\Omega_k y^2)^{1/2} 
\label{H}
\end{equation} 
(e.g. Weinberg 1972),  and 
\begin{equation}
q(z) = -1-(1+z)~y^{\prime \prime} ~(y^{\prime})^{-1} + \Omega_k y y^{\prime}
(1+z)/(1+\Omega_k ~y^2) 
\label{q}
\end{equation} (Daly \& Djorgovski 2003), where 
$H(z)/H_0 \equiv (\dot{a}/a)^2$, $q(z) \equiv - (\ddot{a}a)/(\dot{a})^2$,
$\Omega_k \equiv -k/(H_0a_0)^2$, and $k$ is positive 
and $\Omega_k$ is negative when space
curvature is positive. 
Thus, the zero redshift value of 
$q(z=0)= -1-(y^{\prime \prime}/y^{\prime})\mid_{z=0}$ is 
independent of space curvature, as is $E_0 = (1/y^{\prime})\mid_{z=0}$ 
since $y=0$ at $z=0$.  Thus, the zero redshift values of $E(z)$ and
$q(z)$ obtained from $y^{\prime}$ and $y^{\prime \prime}$ 
are independent of space curvature. And, the zero redshift 
value of $q$ indicates whether the universe is accelerating 
at the current epoch and can be determined independent of $\Omega_k$. 

\subsubsection{Results Obtained with a Sliding Window Fit}

The data and analysis described in section 2.2.1 were used
to obtain  
$H(z)/H_0$ and $q(z)$ using equations \ref{H}
and \ref{q}. The resuls are shown in Figures 
\ref{yEsnprg} and \ref{qsnprg} for the 
combined sample of 222 sources described above. 
The results confirm that the universe
is accelerating at the current epoch.  
For the Davis et al. (2007) sample of 192 supernovae,
we find a zero redshift value of $q$ are 
$q_0(192SN)= -0.48 \pm 0.11$.  For the 
radio galaxies, we find $q_0(30RG)= -0.65 \pm 0.5$, 
consistent with the results obtained using supernovae.
Again, these results depend only upon the 
form of the Robertson-Walker line element and are independent of
space curvature, whether General Relativity is the correct theory
of gravity, and the content or evolution of the contents of the universe.  
For $k=0$, the redshift at which the universe transitions from 
acceleration to deceleration for the Davis et al. (2007)
sample of 192 SN $z_T = 0.77 \pm 0.1$.

We investigated the effect of the size of the window function of 
the value of $q_0$ and the transition redshift for the combined
sample of 192 supernovae and 30 
radio galaxies. For a window function of width
0.6 in redshift, we obtain $q_0 = -0.48 \pm 0.11$, and a transition
redshift $z_T = 0.78 {}^{+0.08}_{-0.27}$; for a window function
of width 0.5 in redshift, we obtain $q_0 = -0.37 \pm 0.13$, and 
a transition redshift $z_T = 0.79 \pm 0.15$; and for a window function
of width 0.4, we obtain $q_0 =  -0.30 \pm 0.18$, and $z_T = 0.71 \pm 0.37$.
These numbers are all consistent, though the uncertainties increase 
as the size of the window function decreases since then fewer points
are used to determine each quantity. 
It is important to increase the number of data points at all redshifts
to that the size of the window function can be decreased.  
These transition redshifts are consistent with those obtained
by Melchiorri, Pagano, \& Pandolfi (2007).  

\subsubsection{Results Obtained With Fits in Independent Redshift Bins}

The data and analysis described in section 2.2.2 were used  
to obtain  
$H(z)/H_0$ and $q(z)$ using equations \ref{H}
and \ref{q}.  The results are shown in Figures 
\ref{binH} and \ref{binq}, and are listed in Table
\ref{binresults}. 

The values of $H(z_{med})$ are
consistent with predictions in the standard LCDM model at about 
one sigma or better. For comparison, predictions obtained in 
a Cardassian model and generalized Chaplygin gas model, described
in section 2.2.2. are also displayed.  
The value of the deceleration parameter at the 
median redshift of the bin, $q(z_{med})$, does not definitely 
show that the universe is accelerating today, using this approach.  
That is, we do not
see that $q$ must be less than zero. If we consider the data split
into two bins, each with 111 data points, and review the results for
the lower redshift bin, we find that at a redshift of 0.02 the
value of $q$ is $-0.28 \pm 0.14$.    
If we consider the data split into three bins, each with 74 data 
points, and review the results for the lowest redshift bin, 
we find that the value $q$ is consistent with zero for all of the
data points in this bin.  Evidently, we do not yet have a sufficient
density of data points even at a redshift less than about 0.5 to 
be able to definitely state at three sigma or better 
that the universe is accelerating today, using this method (fits in independent
redshift bins).  
It is only when we increase the number of data points that
go into the determination of $q(z)$ by using the sliding window
function, that we can conclude that the universe is accelerating today
using our model-independent analysis.  

Of course, when it is concluded that the universe is accelerating
today in the context of a quintessence model or other models, 
all of the data are being used in the context of that particular
model.  The quintessence and most other models 
implicitly assume that our
universe is described by the Robertson-Walker line element,  
the equations of General Relativity apply, 
and that a specific function  
for the redshift evolution of the dark
energy is valid over the redshift interval under study.  
Much more data are needed in order to test independently these 
(perfectly reasonable) assumptions.

\subsection{Determinations of the Properties of the Dark Energy} 

We can solve for the properties of the dark energy as functions
of redshift if a theory of gravity is specified, as shown
by Daly \& Djorgovski (2004).  
Einstein's equations of General Relativity
are used to relate the pressure, $P_{DE}$, energy density,
$\rho_{DE}$, equation of state $w = P_{DE}/\rho_{DE}$, and
potential, $V_{DE}$ and kinetic $K_{DE}$ energy densities
as functions of redshift to the cosmic scale factor 
$a$, and the first and second derivatives of $a$ with respect
to time.  Since the Robertson-Walker line element has been used to relate
the first and second derivatives of $a$ with respect to time
to the first and second derivatives of the coordinate distance
with respect to redshift, $y^{\prime}$
and $y^{\prime \prime}$, the equations of General Relativity can be used 
to solve for the pressure,
energy density, equation of state, and potential and kinetic
energy density of the dark energy in terms of $y^{\prime}$
and $y^{\prime \prime}$, which are obtained directly from 
the data. Here a value of $k=0$ is assumed, and the 
equations of Daly \& Djorgovski (2004) are 
used to solve for the properties of the dark energy
as functions of redshift.  To obtain the pressure,
we only need to specify that Einstein's equations apply. 

\begin{equation}
{P_{DE}(z) \over \rho_{0c}} = 
-(y^{\prime})^{-2}[1+(2/3)(1+z)~y^{\prime \prime}~(y^{\prime})^{-1}]~.
\label{P}
\end{equation}

As pointed out by Daly \& Djorgovski (2004), the zero redshift 
value of $P$ translates into a value of $\Omega_{\Lambda}$ is a
standard LCDM model since in this model $w=-1$, so 
$\Omega_{\Lambda} = \rho_{DE}/\rho_{0c} = -P_{0}/\rho_{0c}$.
Here, this implies that $\Omega_{\Lambda}= 0.64 \pm 0.1$ if
the dark energy is due to a cosmological constant in a universe
with zero space curvature. This value of $\Omega_{\Lambda}$,
which is obtained from the first and second derivatives of the coordinate
distance, is consistent with other measures.

To obtain
the energy density $\rho_{DE}$ of 
the dark energy, $\Omega_m$ must be specified, and
a value of $\Omega_m=0.3$ is adopted here. 
\begin{equation}
{\rho_{DE}(z) \over \rho_{0c}} = (y^{\prime})^{-2}-\Omega_m(1+z)^3~.
\label{f}
\end{equation}

The equation of state $w$ is obtained by taking the ratio
$P_{DE}/\rho_{DE}$ and is given by 
\begin{equation}
w(z) = {-[1+(2/3)(1+z)~y^{\prime \prime}~(y^{\prime})^{-1}] \over 
1-\Omega_m(1+z)^3~(y^{\prime})^{2}}~.
\label{w}
\end{equation}
 
The potential energy density of a dark energy scalar field is 
given by $V = 0.5 (\rho - P)$ so 
\begin{equation}
{V_{DE}(z) \over \rho_{0c}} = (y^{\prime})^{-2}[1+(1+z)y^{\prime \prime}/(3y^{\prime})]-0.5 \Omega_m(1+z)^3 
\label{V}
\end{equation}
and the kinetic energy density is given by $K = 0.5(\rho+P)$ so  
\begin{equation}{K_{DE} \over \rho_{0c}} = 
-y^{\prime \prime}(1+z)/[3(y^{\prime})^{3}]-0.5
\Omega_m(1+z)^3~, 
\label{K}
\end{equation}
since $V = 0.5 (\rho-P)$ and $K=0.5(\rho +P)$.  

We define a new function, the dark energy indicator $s$, which is given
by 
\begin{equation}
s={y^{\prime \prime} \over (y^{\prime})^3 (1+z)^2}~.
\label{s}
\end{equation}
This is motivated by the fact that, for 
zero space curvature,
the predicted value of $s$ is    
\begin{equation}
s_p = -1.5 \Omega_m \left(1+[1+w]{\Omega_{DE}f(z) \over \Omega_m[1+z]^3}\right)
= -1.5 \Omega_m \left(1+[1+w]{\rho_{DE}\over \rho_m} \right)
~\label{sp}
\end{equation}
as indicated by equations (1) and (2); this occurs because  
$y_p^{\prime \prime} \propto (y_p^{\prime})^3(1+z)^2$ when 
$\Omega_k=0$.  Equation \ref{sp} 
is derived assuming only that General Relativity is valid,
space curvature is zero, and there are two mass-energy components
controlling the expansion of the universe, dark energy and
non-relativistic matter including dark matter and baryons.
It is not assumed that the equation of state of the dark energy, $w$,  
is constant, and no functional form for the evolution of the 
energy density of the dark energy, $f(z)$, is assumed.  
If $w=-1$, then $s_p = -1.5 \Omega_m$,
and the value of $s$ provides a new and independent measure of $\Omega_m$.  
Deviations of $s$ from a constant provide 
a measure of the deviations of $w$ from -1; the amount
of the deviation also depends upon the ratio of the energy density
of the dark energy $\rho_{DE}(z)$
to that of the mass density of non-relativistic matter $\rho_m(z)$,
including dark matter and baryons, 
at a given redshift as discussed in more detail below.

\subsubsection{Results Obtained with a sliding window fit}

The analysis described in section 2.2.1 
was 
applied to the combined sample of 192 supernovae from Davis et al. (2007)
and 30 radio galaxies from Daly et al. (2007).   
The values of $y^{\prime}$ and $y^{\prime \prime}$
were substituted into equations \ref{P}, \ref{f}, \ref{w},
\ref{V}, \ref{K}, \ref{s}, and \ref{S} and the results are shown 
in Figures \ref{Pfw}, \ref{KV}, \ref{s}, and \ref{S}.
The
zero redshift values of these parameters are 
$P_0/\rho_{0c} = -0.64 \pm 0.10$, $\rho_{DE}(z=0)/\rho_{0c} = 0.67 \pm 0.05$, 
$w_0 = -0.95 \pm 0.08$, 
$V_0/\rho_{0c} = 0.65 \pm 0.05$, $K_0/\rho_{0c} = 0.01 \pm 0.03$, and 
$s_0 = -0.50 \pm 0.08$.  The zero redshift value of $s$ indicates a value
of $\Omega_m = 0.33 \pm 0.05$ if $w=-1$, and 
the zero redshift value of $P$ indicates
a value of $\Omega_{\Lambda} = 
0.64 \pm 0.10$ if $w=-1$.   
Overall, the results
are consistent with predictions in a standard LCDM model.

As noted above, the dark energy indicator $s$ provides a measure of
whether $w=-1$ over the redshift range shown in Figure \ref{S}.
To illustrate the signature of a value of $w$ that differs from 
$-1$, three curves are overlayed on the figure. Each curve
is obtained using equation \ref{sp} assuming $\Omega_m = 0.3$,
$\Omega_{DE} = 0.7$, $f(z)=1$ and $w$ is constant 
over the redshift range shown, with values of $w$ of 
$-1.2$ (short dashed curve), $w = -1$ (solid line), and
$w=-0.8$ (long dashed curve).  Clearly, the curves with 
$w=-0.8$ and $w=-1.2$ do not provide a good description of the data.
Another way to look at this is that equation \ref{s} and \ref{sp}
indicate that $w=-1-(\rho_m/\rho_{DE})[2s/(3\Omega_m)+1]$ so the 
second part of this is a measure of the deviation of $w$ from 
-1, and at zero redshift the ratio $\rho_m/\rho_{DE} \sim 0.3/0.7$
and $\Omega_m = 0.3$, so $s$ places rather strong constraints on
deviations of $w_0$ from -1.  

Since most of these functions involve combinations of the first
and second derivatives of the coordinate distance, results obtained
with independent redshift bins are quite noisy, so only values
obtained in two independent redshift bins are listed in Table 3.

\subsection{Effect of Space Curvature on $H(z)/H_0$ and $q(z)$}

Space curvature, parameterized by $\Omega_k$, has only a modest effect on 
$q(z)$, as 
illustrated in Figure \ref{kqofz192}.  Positive space curvature,
with negative $\Omega_k$ flattens the $q(z)$ curve and pushes
the redshift at which the universe transitions from 
an accelerating state to a decelerating state to higher redshift.  
This follows since $y^{\prime}$ is known to be positive in our 
universe since the universe is expanding. Similarly, negative
space curvature moves the redshift at which $q=0$ to lower redshift
causing the universe the transition from an accelerating to a 
decelerating state at lower redshift.  
For $k=0$, the redshift at which the universe transitions from 
acceleration to deceleration for the Davis et al. (2007)
sample is $z_T = 0.77 {}^{+0.11}_{-0.24}$.  
For negative space curvature with 
$\Omega_k = 0.1$ this transition redshift is shifted to about 0.6, 
and for positive space curvature with 
$\Omega_k = -0.1$ it shifts to about 0.8.  

Space curvature also affects the shape of 
$H(z)/H_0$ as illustrated in Figure \ref{kEofz192}. 
Positive values of $\Omega_k$ cause $H(z)$ to 
increase more steeply with redshift than negative values
of $\Omega_k$, which tend to flatten out the $H(z)$ curve.

Thus, the effect of space curvature on $H(z)$ and $q(z)$ is small relative to
the overall uncertainties of their measurements, within a plausible range of
the curvature parameter $\Omega_k$. 

\subsection{Determinations of $y^{\prime}$ and $y^{\prime \prime}$ for other samples}

We also consider the full sample of 260 sources including the 
38 SZ clusters of Bonamente et al. (2006), the 192 supernovae
of Davis et al. (2007), and the 30 radio galaxies of Daly et al.
(2007).  These results are shown in Figure \ref{yetal260}; a window
function of width 0.6 was used to analyze the data.  
This analysis illustrates how this method can be applied to 
diverse data sets.  However, while the SZ cluster distances currently
provide a useful tool for measurements of the far-field Hubble parameter,
it is probably premature to use them as standard rulers to probe the
global geometry and kinematics of the universe at this time.

Gamma-ray bursts as standard candles
have been analyzed in detail by Schaefer (2007),
who gives other pertinent references.
The values of $\mu$ listed by Schaefer (2007) can be used to
determine the dimensionless coordinate distance to each source 
if the value of $H_0$ relevant for the sample is known,
as described in section 2.1, 
and a value of $H_0 = 70 \hbox{km s}^{-1} \hbox{ Mpc}^{-1}$
was used (Schaefer, private communication).  
The dimensionless coordinate distances were analyzed
to determine the functions $y(z)$, $y^{\prime}(z)$,
and $y^{\prime \prime}(z)$ using a window function of
width 2.0 in redshift. 
A first look at results for the gamma-ray bursts are 
shown in Fig. 
\ref{GRB}, which
suggest that these are a potentially promising 
tool to study cosmology at very large distances, and are
broadly consistent with predictions in the canonical LCDM
model, but the quality and the sparsity of the data are still not
sufficient for the model-independent analysis as shown above for the
SN+RG sample.

\section{SUMMARY AND CONCLUSIONS}

The work presented here improves and extends our previous results. 
First, expanded and improved data sets are considered:
three supernova samples and one radio galaxy sample.
The radio galaxy
data set has 11 new sources, increasing its size to 30 sources,
and the supernovae data sets have increased substantially in 
size and quality.  In addition, SZ cluster distances and gamma-ray 
burst distances are considered.  The dimensionless coordinate distances
(obtained directly from the data), and first and second 
derivatives of the 
distance are obtained as functions of redshift using a sliding
window fit.  The good agreement obtained using
supernovae and radio galaxies, two completely independent methods,
with sources that cover similar redshift ranges, suggests that neither
method is strongly affected by systematic effects, and that 
each method provides a reliable cosmological tool.

The first and second derivatives of the distance are combined to
obtain the acceleration parameter $q(z)$, allowing for non-zero
space curvature. It is shown that the zero redshift
value of $q(z)$, $q_o$, is independent of space curvature, and can
be obtained from the first and second derivatives of the
coordinate distance.  Thus, $q_0$, which indicates whether the universe
is accelerating at the current epoch, can be obtained directly from the 
supernova and radio galaxy data; our determinations of $q(z)$ 
only relies upon
the validity of the Robertson-Walker line element, and is
independent of a theory of gravity, and the contents of the universe.  
Each of the supernova samples, analyzed using a sliding window fit,
indicate
that the universe is accelerating today independent of space curvature,
independent of whether General Relativity is the correct theory of gravity,
and of the contents of the universe. The effect of non-zero space
curvature on $q(z)$ is to shift the redshift at which the universe
transitions from acceleration to deceleration, moving this to lower
redshift for negative space curvature and to higher redshift for 
positive space. The zero redshift values of $q$ obtained 
using a sliding window fit. 
for
the Davis et al. (2007) supernova sample is 
$q_0(192SN)= -0.48 \pm 0.11$
and that obtained for the radio galaxy sample of 
Daly et al. (2007) is $q_0(30RG)=  -0.65 \pm 0.53$ indicating
that the universe is accelerating at the current epoch.
The data were also binned so that only certain subsets of the
data were used to solve for $y^{\prime}$, $y^{\prime \prime}$,
$H(z)/H_0$, and $q(z)$.  The results for $y^{\prime}$ and 
$H(z)/H_0$ indicate that the standard LCDM model provides a 
good description of the data.  The results for $y^{\prime \prime}$
and $q(z)$ are consistent with the standard LCDM model, but
do not independently confirm the model or the acceleration
of the universe.

In addition to the evaluation of the standard cosmological parameters,
in an even more direct approach, we compared $y^{\prime}$
and $y^{\prime \prime}$ obtained from the fits to the data to model predictions.
Comparisons of $y^{\prime}$
and $y^{\prime \prime}$ with predictions based on General Relativity
indicate that General Relativity provides an accurate description of
the data on look-back time scales of about ten billion years, thus providing
a very large scale test of General Relativity.

Another new approach is that the data were analyzed using both a sliding window fit
and fits in independent redshift bins.  The fits in statistically
independent redshift bins are broadly consistent with the sliding window fits,
but are generally noisier (as expected).
 
We also explored the effects of non-zero space curvature
on determinations of $H(z)$ and $q(z)$.  It is shown that
the zero redshift value of $q$, obtained by applying equation (4)
to $y^{\prime}$ and $y^{\prime \prime}$, is independent of space curvature.
This means that our method can be used to determine $q_0$, and thus
the degree to which the universe is accelerating at the current epoch, with 
only one assumption, that the Robertson-Walker
line element is valid. In addition, it is found that 
the effect of space curvature on
the shape of $H(z)$ and $q(z)$ is small, relative to the uncertainties
arising from the measurement errors.  

After determining the expansion and acceleration rates of the
universe as functions of redshift independent of a theory
of gravity, we solve for the pressure, energy density, equation
of state, and potential and kinetic energy of the dark energy
as functions of redshift assuming that General Relativity is
the correct theory of gravity.  We also define a new function,
the dark energy indicator $s$, which provides a measure of
deviations of the equation of state of the dark energy 
$w$ from $-1$, and provides a new and independent measure of
$\Omega_m$ if $w=-1$.  
The results obtained using a sliding window fit indicate
that a cosmological constant in 
a spatially flat universe provides a good description of
each of these quantities over the redshift range from zero to 
one. The zero redshift values of these quantities 
obtained with the Davis et al. (2007) supernovae sample
are
$P_{DE,0}/\rho_{0c} = -0.64 \pm 0.10$, $\rho_{DE,0}/\rho_{0c} 
= 0.67 \pm 0.05$, 
$w_{DE,0} = -0.95 \pm 0.08$, 
$V_{DE,0}/\rho_{0c} = 0.65 \pm 0.05$, 
$K_{DE,0}/\rho_{0c} = 0.01 \pm 0.03$, and 
$s_0 = -0.50 \pm 0.08$. 
In the standard Lambda-Cold Dark Matter Model, 
$\Omega_{\Lambda} = - P_0/\rho_{0c} = 0.64 \pm 0.1$, 
obtained using the first and second derivatives of the coordinate distance, 
provides an independent measure
of $\Omega_{\Lambda}$.  In addition, in this model, 
$w=-1$, so $s$ provides a measure of $\Omega_m$, and 
the value obtained here using the first and second derivatives
of the coordinate distance, is $\Omega_m = 0.33 \pm 0.05$. 
Overall, the shapes of the pressure, energy density,
equation of state, and other parameters as functions 
are redshift are consistent
with those predicted in a standard LCDM model.  There is a tantalizing
hint that there may be divations from the standard model at high
redshift; more observations at high redshift will be needed to investigate
this further. The results obtained using fits in independent redshift bins
are consistent with the standard LCDM model, but
do not independently confirm the model.

\acknowledgements
We would like to thank the observers for their tireless efforts in 
obtaining the data used for this study. We would also like to thank
the referee for helpful comments and suggesitons.
This work was supported in part by U. S. National Science
Foundation grants AST-0507465 (R.A.D.) and AST-0407448 (S.G.D.),
and the Ajax Foundation (S.G.D.).

\begin{deluxetable}{lllll}
\tablewidth{0pt}
\tablecaption{Distances 192 Supernovae, 30 Radio Galaxies, and 38 Galaxy Clusters}
\tablehead{
\colhead{Type} & \colhead{Source} & \colhead{$z$} &  \colhead{$y$} & \colhead{$\sigma_y$} 
 }
\startdata  
SN	&	sn1994S	&	0.016	&	0.0160	&	0.0016	\\
SN	&	sn2001V	&	0.016	&	0.0145	&	0.0015	\\
SN	&	sn1996bo	&	0.016	&	0.0135	&	0.0015	\\
SN	&	sn2001cz	&	0.016	&	0.0155	&	0.0017	\\
SN	&	sn2000dk	&	0.016	&	0.0161	&	0.0016	\\
SN	&	sn1997Y	&	0.017	&	0.0174	&	0.0018	\\
SN	&	sn1998ef	&	0.017	&	0.0146	&	0.0016	\\
SN	&	sn1998V	&	0.017	&	0.0160	&	0.0016	\\
SN	&	sn1999ek	&	0.018	&	0.0155	&	0.0016	\\
SN	&	sn1992bo	&	0.018	&	0.0190	&	0.0018	\\
SN	&	sn1992bc	&	0.020	&	0.0200	&	0.0017	\\
SN	&	sn2000fa	&	0.022	&	0.0205	&	0.0020	\\
SN	&	sn1995ak	&	0.022	&	0.0187	&	0.0018	\\
SN	&	sn2000cn	&	0.023	&	0.0226	&	0.0019	\\
SN	&	sn1998eg	&	0.024	&	0.0248	&	0.0023	\\
SN	&	sn1994M	&	0.024	&	0.0239	&	0.0022	\\
SN	&	sn2000ca	&	0.025	&	0.0239	&	0.0019	\\
SN	&	sn1993H	&	0.025	&	0.0224	&	0.0019	\\
SN	&	sn1992ag	&	0.026	&	0.0228	&	0.0023	\\
SN	&	sn1999gp	&	0.026	&	0.0284	&	0.0021	\\
SN	&	sn1992P	&	0.026	&	0.0281	&	0.0025	\\
SN	&	sn1996C	&	0.028	&	0.0329	&	0.0027	\\
SN	&	sn1998ab	&	0.028	&	0.0231	&	0.0019	\\
SN	&	sn1997dg	&	0.030	&	0.0361	&	0.0032	\\
SN	&	sn2001ba	&	0.031	&	0.0319	&	0.0023	\\
SN	&	sn1990O	&	0.031	&	0.0309	&	0.0024	\\
SN	&	sn1999cc	&	0.032	&	0.0310	&	0.0024	\\
SN	&	sn1996bl	&	0.035	&	0.0350	&	0.0029	\\
SN	&	sn1994T	&	0.036	&	0.0338	&	0.0026	\\
SN	&	sn2000cf	&	0.037	&	0.0395	&	0.0031	\\
SN	&	sn1999aw	&	0.039	&	0.0429	&	0.0026	\\
SN	&	sn1992bl	&	0.043	&	0.0417	&	0.0035	\\
SN	&	sn1992bh	&	0.045	&	0.0505	&	0.0044	\\
SN	&	sn1995ac	&	0.049	&	0.0431	&	0.0032	\\
SN	&	sn1993ag	&	0.050	&	0.0541	&	0.0045	\\
SN	&	sn1990af	&	0.050	&	0.0454	&	0.0042	\\
SN	&	sn1993O	&	0.052	&	0.0553	&	0.0038	\\
SN	&	sn1998dx	&	0.054	&	0.0503	&	0.0035	\\
RG	&	3C 405	&	0.056	&	0.0514	&	0.0105	\\
SN	&	sn1992bs	&	0.063	&	0.0695	&	0.0064	\\
SN	&	sn1993B	&	0.071	&	0.0736	&	0.0064	\\
SN	&	sn1992ae	&	0.075	&	0.0713	&	0.0069	\\
SN	&	sn1992bp	&	0.079	&	0.0731	&	0.0050	\\
SN	&	sn1992br	&	0.088	&	0.0718	&	0.0076	\\
SN	&	sn1992aq	&	0.101	&	0.1145	&	0.0079	\\
SN	&	sn1996ab	&	0.124	&	0.1174	&	0.0108	\\
CL	&	Abell~1413	&	0.142	&	0.2283	&	0.0454	\\
CL	&	Abell~2204 	&	0.152	&	0.1801	&	0.0192	\\
SN	&	e020	&	0.159	&	0.1716	&	0.0229	\\
CL	&	Abell~2259	&	0.164	&	0.1731	&	0.0806	\\
CL	&	Abell~586 	&	0.171	&	0.1561	&	0.0405	\\
CL	&	Abell~1914 	&	0.171	&	0.1321	&	0.0135	\\
CL	&	Abell~2218	&	0.176	&	0.1990	&	0.0377	\\
SN	&	k429	&	0.181	&	0.1764	&	0.0138	\\
CL	&	Abell~665 	&	0.182	&	0.2000	&	0.0288	\\
CL	&	Abell~1689	&	0.183	&	0.1971	&	0.0273	\\
CL	&	Abell~2163	&	0.202	&	0.1602	&	0.0139	\\
SN	&	d086	&	0.205	&	0.1887	&	0.0261	\\
SN	&	h363	&	0.213	&	0.2103	&	0.0320	\\
SN	&	n404	&	0.216	&	0.2364	&	0.0338	\\
CL	&	Abell~773 	&	0.217	&	0.3057	&	0.0484	\\
SN	&	g005	&	0.218	&	0.2133	&	0.0255	\\
CL	&	Abell~2261	&	0.224	&	0.2290	&	0.0518	\\
CL	&	Abell~2111	&	0.229	&	0.2016	&	0.0583	\\
CL	&	Abell~267 	&	0.230	&	0.1892	&	0.0315	\\
CL	&	RX~J2129.7+0005	&	0.235	&	0.1456	&	0.0301	\\
SN	&	e132	&	0.239	&	0.2146	&	0.0287	\\
CL	&	Abell~1835	&	0.252	&	0.3434	&	0.0160	\\
CL	&	Abell~68 	&	0.255	&	0.2027	&	0.0563	\\
SN	&	04D3ez	&	0.263	&	0.2462	&	0.0238	\\
SN	&	n326	&	0.268	&	0.2509	&	0.0300	\\
SN	&	k425	&	0.274	&	0.2881	&	0.0371	\\
CL	&	Abell~697	&	0.282	&	0.2892	&	0.0871	\\
SN	&	p455	&	0.284	&	0.2832	&	0.0378	\\
SN	&	03D4ag	&	0.285	&	0.2678	&	0.0160	\\
SN	&	m027	&	0.286	&	0.3447	&	0.0508	\\
CL	&	Abell~611 	&	0.288	&	0.2575	&	0.0594	\\
CL	&	ZW~3146	&	0.291	&	0.2747	&	0.0066	\\
SN	&	03D3ba	&	0.291	&	0.2196	&	0.0324	\\
SN	&	g055	&	0.302	&	0.3192	&	0.0544	\\
SN	&	d117	&	0.309	&	0.3219	&	0.0400	\\
SN	&	n278	&	0.309	&	0.2856	&	0.0276	\\
CL	&	Abell~1995	&	0.322	&	0.4033	&	0.0491	\\
CL	&	MS~1358.4+6245	&	0.327	&	0.3844	&	0.0323	\\
SN	&	03D1fc	&	0.331	&	0.2996	&	0.0290	\\
SN	&	e029	&	0.332	&	0.3297	&	0.0425	\\
SN	&	d083	&	0.333	&	0.2279	&	0.0147	\\
SN	&	04D3kr	&	0.337	&	0.3209	&	0.0251	\\
SN	&	g097	&	0.340	&	0.3354	&	0.0479	\\
SN	&	04D3nh	&	0.340	&	0.3463	&	0.0271	\\
SN	&	m193	&	0.341	&	0.2960	&	0.0313	\\
SN	&	d149	&	0.342	&	0.3459	&	0.0334	\\
SN	&	h364	&	0.344	&	0.2994	&	0.0234	\\
SN	&	03D1bp	&	0.346	&	0.3248	&	0.0374	\\
SN	&	h359	&	0.348	&	0.3881	&	0.0483	\\
SN	&	e136	&	0.352	&	0.3417	&	0.0425	\\
SN	&	04D2fs	&	0.357	&	0.3452	&	0.0350	\\
SN	&	04D3fk	&	0.358	&	0.3089	&	0.0299	\\
SN	&	d093	&	0.363	&	0.3566	&	0.0230	\\
SN	&	n263	&	0.368	&	0.3285	&	0.0257	\\
SN	&	03D3ay	&	0.371	&	0.3662	&	0.0388	\\
CL	&	Abell~370  	&	0.375	&	0.3807	&	0.0687	\\
SN	&	g052	&	0.383	&	0.3250	&	0.0329	\\
SN	&	g142	&	0.399	&	0.3862	&	0.0765	\\
SN	&	d085	&	0.401	&	0.3857	&	0.0391	\\
SN	&	k448	&	0.401	&	0.4594	&	0.0846	\\
RG	&	3C142.1	&	0.406	&	0.3325	&	0.0607	\\
CL	&	MACS~J2228.5+2036	&	0.412	&	0.4416	&	0.0851	\\
SN	&	04D2fp	&	0.415	&	0.3999	&	0.0313	\\
SN	&	k485	&	0.416	&	0.4184	&	0.0751	\\
SN	&	g133	&	0.421	&	0.4286	&	0.0651	\\
SN	&	h342	&	0.421	&	0.4208	&	0.0310	\\
SN	&	f235	&	0.422	&	0.3498	&	0.0387	\\
SN	&	b013	&	0.426	&	0.3824	&	0.0405	\\
SN	&	e148	&	0.429	&	0.4321	&	0.0398	\\
SN	&	04D2gb	&	0.430	&	0.3526	&	0.0292	\\
RG	&	3C 244.1	&	0.430	&	0.3631	&	0.0671	\\
SN	&	d089	&	0.436	&	0.3922	&	0.0361	\\
SN	&	d097	&	0.436	&	0.4013	&	0.0314	\\
SN	&	03D3aw	&	0.449	&	0.3923	&	0.0434	\\
CL	&	RX~J1347.5-1145	&	0.451	&	0.3571	&	0.0260	\\
SN	&	04D3gt	&	0.451	&	0.2812	&	0.0298	\\
SN	&	HST04Yow	&	0.460	&	0.4114	&	0.0625	\\
SN	&	03D3cd	&	0.461	&	0.3982	&	0.0330	\\
SN	&	03D3cc	&	0.463	&	0.4222	&	0.0331	\\
SN	&	m158	&	0.463	&	0.4914	&	0.0634	\\
SN	&	e108	&	0.469	&	0.4262	&	0.0314	\\
SN	&	04D3df	&	0.470	&	0.3796	&	0.0350	\\
SN	&	sn2002dc	&	0.475	&	0.4091	&	0.0396	\\
CL	&	MACS~J2214.9-1359	&	0.483	&	0.5474	&	0.0950	\\
CL	&	MACS~J1311.0-0310 	&	0.490	&	0.5271	&	0.1604	\\
SN	&	g160	&	0.493	&	0.4391	&	0.0526	\\
SN	&	h319	&	0.495	&	0.4426	&	0.0428	\\
SN	&	03D1ax	&	0.496	&	0.4283	&	0.0394	\\
SN	&	e149	&	0.497	&	0.4087	&	0.0489	\\
SN	&	h283	&	0.502	&	0.4592	&	0.0782	\\
SN	&	03D1au	&	0.504	&	0.4713	&	0.0391	\\
SN	&	p524	&	0.508	&	0.4449	&	0.0451	\\
SN	&	g120	&	0.510	&	0.4185	&	0.0405	\\
SN	&	d084	&	0.519	&	0.5612	&	0.0749	\\
RG	&	3C172	&	0.519	&	0.6635	&	0.1366	\\
SN	&	04D2gc	&	0.521	&	0.4431	&	0.0673	\\
SN	&	04D1ak	&	0.526	&	0.4520	&	0.0604	\\
SN	&	n285	&	0.528	&	0.4814	&	0.0576	\\
SN	&	d033	&	0.531	&	0.5594	&	0.0438	\\
SN	&	03D3af	&	0.532	&	0.5289	&	0.0706	\\
SN	&	f011	&	0.539	&	0.4846	&	0.0558	\\
SN	&	f244	&	0.540	&	0.4979	&	0.0596	\\
CL	&	CL~0016+1609 	&	0.541	&	0.5451	&	0.0869	\\
CL	&	MACS~J1149.5+2223 	&	0.544	&	0.3166	&	0.0693	\\
CL	&	MACS~J1423.8+2404	&	0.545	&	0.5901	&	0.0178	\\
RG	&	3C 330	&	0.549	&	0.3424	&	0.0652	\\
CL	&	MS~0451.6-0305 	&	0.550	&	0.5642	&	0.0973	\\
SN	&	04D4bq	&	0.550	&	0.5016	&	0.0670	\\
SN	&	04D3hn	&	0.552	&	0.4035	&	0.0762	\\
SN	&	f041	&	0.561	&	0.4912	&	0.0385	\\
CL	&	MACS~J2129.4-0741	&	0.570	&	0.5352	&	0.1308	\\
SN	&	03D4gf	&	0.581	&	0.5149	&	0.0427	\\
SN	&	03D1aw	&	0.582	&	0.5828	&	0.0564	\\
CL	&	MS~2053.7-0449	&	0.583	&	1.0063	&	0.1725	\\
CL	&	MACS~J0647.7+7015 	&	0.584	&	0.3126	&	0.0792	\\
SN	&	03D4gg	&	0.592	&	0.5161	&	0.0594	\\
SN	&	h323	&	0.603	&	0.5467	&	0.0554	\\
SN	&	03D4dy	&	0.604	&	0.4937	&	0.0728	\\
SN	&	04D3do	&	0.610	&	0.4987	&	0.0666	\\
SN	&	e138	&	0.612	&	0.5386	&	0.0446	\\
SN	&	04D4an	&	0.613	&	0.5611	&	0.1008	\\
SN	&	f231	&	0.619	&	0.5513	&	0.0432	\\
SN	&	04D3co	&	0.620	&	0.5904	&	0.0707	\\
SN	&	03D4dh	&	0.627	&	0.5216	&	0.0552	\\
RG	&	3C 337	&	0.630	&	0.5064	&	0.0704	\\
SN	&	e140	&	0.631	&	0.5084	&	0.0421	\\
SN	&	n256	&	0.631	&	0.5575	&	0.0385	\\
SN	&	g050	&	0.633	&	0.4805	&	0.0398	\\
SN	&	03D4at	&	0.633	&	0.6021	&	0.0693	\\
RG	&	3C169.1	&	0.633	&	0.6222	&	0.0708	\\
SN	&	sn2003be	&	0.640	&	0.5246	&	0.0628	\\
SN	&	04D3cy	&	0.643	&	0.6209	&	0.0629	\\
SN	&	e147	&	0.645	&	0.5327	&	0.0442	\\
RG	&	3C44	&	0.660	&	0.7621	&	0.0848	\\
SN	&	sn2003bd	&	0.670	&	0.5597	&	0.0644	\\
SN	&	03D1co	&	0.679	&	0.6817	&	0.0848	\\
CL	&	MACS~J0744.8+3927 	&	0.686	&	0.7261	&	0.1858	\\
SN	&	g240	&	0.687	&	0.5267	&	0.0485	\\
SN	&	h300	&	0.687	&	0.5390	&	0.0422	\\
SN	&	03D1fl	&	0.688	&	0.5486	&	0.0581	\\
RG	&	3C34	&	0.690	&	0.5920	&	0.0644	\\
SN	&	04D2iu	&	0.691	&	0.6259	&	0.1124	\\
SN	&	p454	&	0.695	&	0.6569	&	0.0514	\\
SN	&	03D4cz	&	0.695	&	0.5414	&	0.0848	\\
RG	&	3C441	&	0.707	&	0.5340	&	0.0667	\\
SN	&	04D3is	&	0.710	&	0.7074	&	0.1108	\\
RG	&	3C 55	&	0.720	&	0.5852	&	0.0777	\\
SN	&	04D1aj	&	0.721	&	0.6264	&	0.0664	\\
SN	&	04D3fq	&	0.730	&	0.6556	&	0.0785	\\
SN	&	sn2002kd	&	0.735	&	0.5265	&	0.0485	\\
SN	&	HST04Rak	&	0.740	&	0.5863	&	0.0621	\\
SN	&	04D2ja	&	0.741	&	0.6544	&	0.0723	\\
RG	&	3C 247	&	0.749	&	0.5430	&	0.0682	\\
SN	&	04D3ks	&	0.752	&	0.5877	&	0.0622	\\
CL	&	MS~1137.5+6625 	&	0.784	&	1.3033	&	0.2629	\\
SN	&	03D4fd	&	0.791	&	0.6880	&	0.0665	\\
RG	&	3C41	&	0.794	&	0.6315	&	0.0714	\\
SN	&	03D1fq	&	0.800	&	0.7403	&	0.0920	\\
RG	&	3C 265	&	0.811	&	0.5901	&	0.0793	\\
CL	&	RX~J1716.4+6708	&	0.813	&	0.4833	&	0.2184	\\
RG	&	3C114	&	0.815	&	0.6368	&	0.0695	\\
SN	&	04D3nc	&	0.817	&	0.6688	&	0.0647	\\
SN	&	03D4cn	&	0.818	&	0.8111	&	0.1009	\\
SN	&	04D3lu	&	0.822	&	0.6795	&	0.0688	\\
CL	&	MS~1054.5-0321	&	0.826	&	0.6225	&	0.1264	\\
RG	&	3C54	&	0.827	&	0.7573	&	0.0835	\\
SN	&	04D3cp	&	0.830	&	0.6342	&	0.0496	\\
SN	&	HST05Spo	&	0.839	&	0.5729	&	0.0554	\\
SN	&	04D4bk	&	0.840	&	0.7110	&	0.0557	\\
SN	&	sn2003eq	&	0.840	&	0.6336	&	0.0642	\\
RG	&	3C6.1	&	0.840	&	0.7402	&	0.0861	\\
SN	&	HST04Man	&	0.854	&	0.7187	&	0.0993	\\
RG	&	3C 325	&	0.860	&	0.7080	&	0.1292	\\
SN	&	03D1ew	&	0.868	&	0.7233	&	0.0566	\\
CL	&	CL~J1226.9+3332	&	0.890	&	0.5232	&	0.1696	\\
SN	&	sn2003eb	&	0.900	&	0.6052	&	0.0725	\\
SN	&	03D4di	&	0.905	&	0.6742	&	0.0497	\\
SN	&	04D3gx	&	0.910	&	0.7973	&	0.0661	\\
SN	&	04D3ki	&	0.930	&	0.8732	&	0.0764	\\
SN	&	sn2003XX	&	0.935	&	0.6918	&	0.0956	\\
SN	&	03D4cx	&	0.949	&	0.7996	&	0.0626	\\
SN	&	04D3ml	&	0.950	&	0.7562	&	0.0522	\\
SN	&	sn2002dd	&	0.950	&	0.6897	&	0.1112	\\
SN	&	sn2003es	&	0.954	&	0.7975	&	0.1028	\\
SN	&	HST04Tha	&	0.954	&	0.6482	&	0.0836	\\
RG	&	3C 289	&	0.967	&	0.5950	&	0.1097	\\
SN	&	HST04Pat	&	0.970	&	0.9380	&	0.1598	\\
RG	&	3C 268.1	&	0.974	&	0.7458	&	0.1432	\\
SN	&	HST04Omb	&	0.975	&	0.7570	&	0.0941	\\
RG	&	3C 280	&	0.996	&	0.6477	&	0.1221	\\
SN	&	04D3dd	&	1.010	&	0.9494	&	0.0743	\\
SN	&	HST05Str	&	1.010	&	0.9627	&	0.0887	\\
SN	&	HST05Fer	&	1.020	&	0.6688	&	0.0862	\\
SN	&	HST04Eag	&	1.020	&	0.8537	&	0.0786	\\
RG	&	3C 356	&	1.079	&	0.8735	&	0.1842	\\
SN	&	HST05Gab	&	1.120	&	0.8716	&	0.0763	\\
SN	&	sn2002ki	&	1.140	&	0.8795	&	0.1215	\\
SN	&	HST04Gre	&	1.140	&	0.7767	&	0.1145	\\
RG	&	3C 267	&	1.144	&	0.7136	&	0.1396	\\
RG	&	3C 194	&	1.190	&	1.0074	&	0.2047	\\
RG	&	3C 324	&	1.210	&	1.0208	&	0.3089	\\
SN	&	HST05Koe	&	1.230	&	1.0432	&	0.1153	\\
SN	&	HST05Lan	&	1.230	&	0.9514	&	0.0920	\\
SN	&	sn2002fw	&	1.300	&	0.9615	&	0.0930	\\
SN	&	sn2002hp	&	1.305	&	0.7447	&	0.1063	\\
RG	&	3C469.1	&	1.336	&	1.1364	&	0.3255	\\
SN	&	sn2003dy	&	1.340	&	0.8860	&	0.1306	\\
SN	&	HST04Mcg	&	1.370	&	1.0091	&	0.1208	\\
SN	&	HST04Sas	&	1.390	&	0.8595	&	0.0792	\\
RG	&	3C 437	&	1.480	&	0.9299	&	0.2697	\\
RG	&	3C 68.2	&	1.575	&	1.5748	&	0.4873	\\
RG	&	3C 322	&	1.681	&	1.3078	&	0.3400	\\
SN	&	sn1977ff	&	1.755	&	0.9174	&	0.1521	\\
RG	&	3C 239	&	1.790	&	1.3666	&	0.3382	\\
\enddata
\label{yTable}
\end{deluxetable}

\begin{deluxetable}{lllllllll}
\tablewidth{0pt}
\tablecaption{Fits in Independent Redshift Bins to 192 Supernovae and 30 Radio Galaxies}
\tablehead{
\colhead{Bin} &\colhead{N} & \colhead{$z_{med}$} & \colhead{$z_{min}$} &  
\colhead{$z_{max}$} &\colhead{$y^{\prime}(z_{med})$} 
& \colhead{$H(z_{med})/H_0$} & 
\colhead{$y^{\prime \prime}(z_{med})$} & \colhead{$q(z_{med})$}    }
\startdata 
1/6&37	&	0.025	&	0.016	&	0.052	&$	1.09	\pm	0.05	$&$	0.92	\pm	0.05	$&$	-6.2	\pm	7.5	$&$	4.8	\pm	6.9	$\\
2/6&37	&	0.275	&	0.054	&	0.348	&$	0.74	\pm	0.07	$&$	1.34	\pm	0.13	$&$	-1.6	\pm	0.8	$&$	1.9	\pm	1.6	$\\
3/6&37	&	0.430	&	0.352	&	0.504	&$	0.67	\pm	0.15	$&$	1.48	\pm	0.32	$&$	2.7	\pm	6.50	$&---	\\			
4/6&37	&	0.600	&	0.508	&	0.670	&$	0.72	\pm	0.28	$&$	1.4	\pm	0.5	$&$	1.2	\pm	11	$&---	\\			
5/6&37	&	0.790	&	0.679	&	0.905	&$	0.55	\pm	0.16	$&$	1.8	\pm	0.5	$&$	-4.0	\pm	5.0	$&---	\\			
6/6&37	&	1.100	&	0.910	&	1.790	&$	0.40	\pm	0.13	$&$	2.5	\pm	0.8	$&$	-0.4	\pm	0.8	$&$	1.1	\pm	3.7	$\\
\\																							
1/4&55	&	0.035	&	0.016	&	0.268	&$	1.03	\pm	0.03	$&$	0.97	\pm	0.03	$&$	-0.9	\pm	0.6	$&$	-0.1	\pm	0.6	$\\
2/4&55	&	0.400	&	0.274	&	0.502	&$	0.79	\pm	0.08	$&$	1.27	\pm	0.13	$&$	-0.4	\pm	2.5	$&$	-0.2	\pm	4.5	$\\
3/4&55	&	0.630	&	0.504	&	0.749	&$	0.54	\pm	0.12	$&$	1.85	\pm	0.40	$&$	-2.50	\pm	3.2	$&---	\\			
4/4&57	&	0.965	&	0.752	&	1.790	&$	0.64	\pm	0.10	$&$	1.57	\pm	0.23	$&$	-0.9	\pm	0.5	$&$	1.80	\pm	1.3	$\\
\\																							
																							
1/3&74	&	0.050	&	0.016	&	0.348	&$	1.02	\pm	0.03	$&$	0.98	\pm	0.03	$&$	-1.05	\pm	0.27	$&$	0.08	\pm	0.26	$\\
2/3&74	&	0.505	&	0.352	&	0.670	&$	0.80	\pm	0.06	$&$	1.26	\pm	0.09	$&$	0.9	\pm	1.4	$&$	-2.7	\pm	2.6	$\\
3/3&74	&	0.905	&	0.679	&	1.790	&$	0.66	\pm	0.06	$&$	1.52	\pm	0.15	$&$	-0.8	\pm	0.3	$&$	1.2	\pm	0.9	$\\
\\																							
1/2&111	&	0.275	&	0.016	&	0.504	&$	0.84	\pm	0.02	$&$	1.19	\pm	0.03	$&$	-0.73	\pm	0.16	$&$	0.10	\pm	0.27	$\\
2/2&111	&	0.790	&	0.508	&	1.790	&$	0.63	\pm	0.04	$&$	1.59	\pm	0.10	$&$	-0.40	\pm	0.21	$&$	0.13	\pm	0.57	$\\

\enddata
\label{binresults}
\end{deluxetable}

\begin{deluxetable}{lllllll}
\tablewidth{0pt}
\tablecaption{Fits in Independent Redshift Bins at the Median Redshift of the Bin}
\tablehead{
\colhead{Data}& 
\colhead{$P_{DE}/\rho_{0c}$} & \colhead{$\rho_{DE}/\rho_{0c}$} & \colhead{$w_{DE}$} & \colhead{$V_{DE}/\rho_{0c}$} &
\colhead{$K_{DE}/\rho_{0c}$} & \colhead{$s $} }
\startdata 
										1/2&$	-0.38	\pm	0.24	$&$	0.79	\pm	0.07	$&$	-0.48	\pm	0.34	$&$	0.58	\pm	0.10	$&$	0.21	\pm	0.15	$&$-0.75 \pm  0.21$\\
2/2&$	-0.6	\pm	1.0	$&$	0.8	\pm	0.3	$&$	-0.8	\pm	1.2	$&$	0.7	\pm	0.6	$&$	0.1	\pm	0.45	$&$ -0.50 \pm  0.25$\\

\enddata
\label{morebinresults}
\end{deluxetable}

\begin{figure}
\plotone{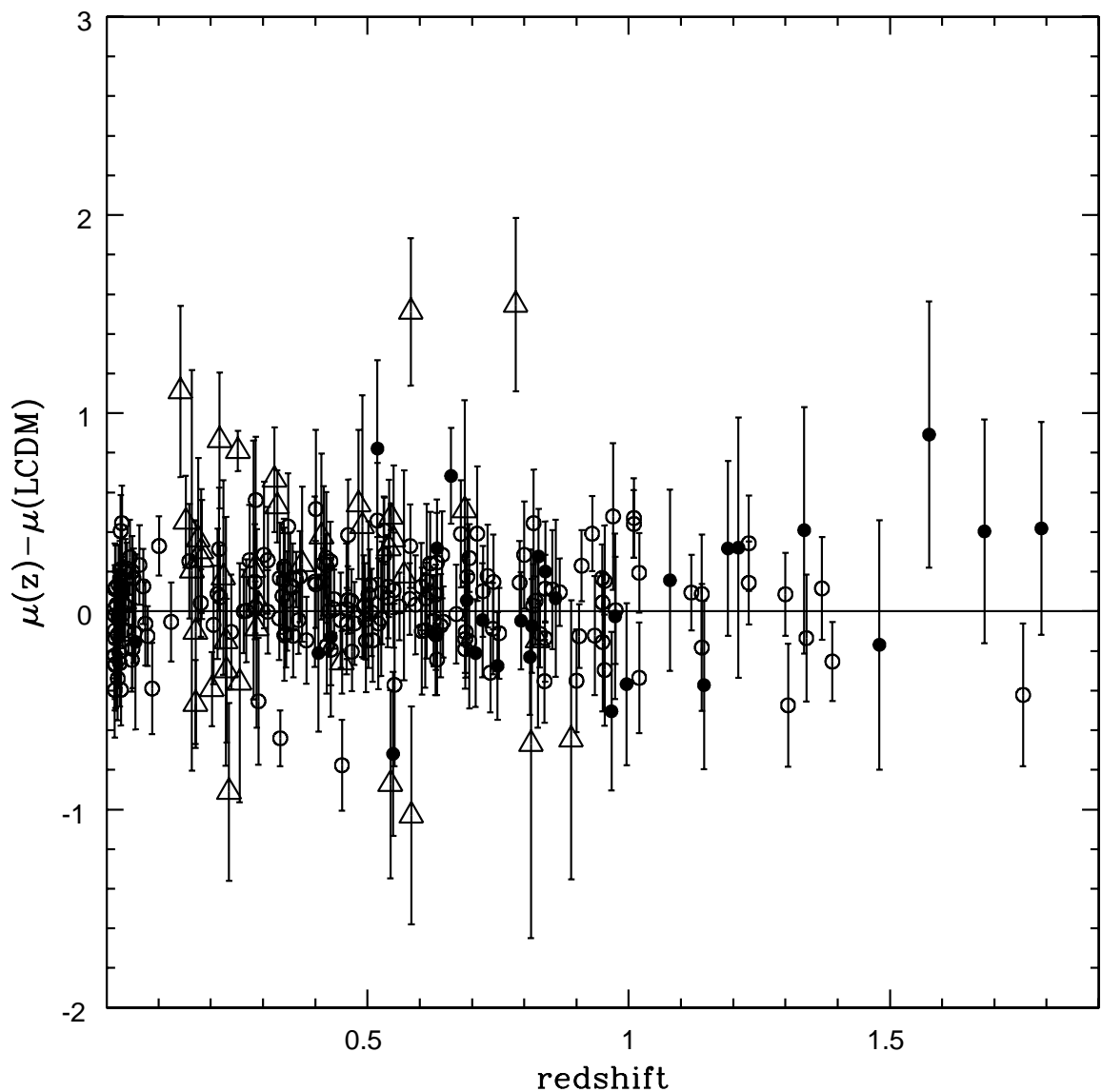}
\caption{The difference between the distance modulus to the source and 
that expected in a standard LCDM model with 
$\Omega_m = 0.3$ and $\Omega_{\Lambda}=0.7$.  Open circles represent
the 192 supernovae of Davis et al. (2007), filled circles
represent the 30 radio galaxies of Daly et al. (2007), and 
open triangles represent the 38 clusters of Bonamente et al. (2006). }
\label{mudiffofz}
\end{figure}

\begin{figure}
\plotone{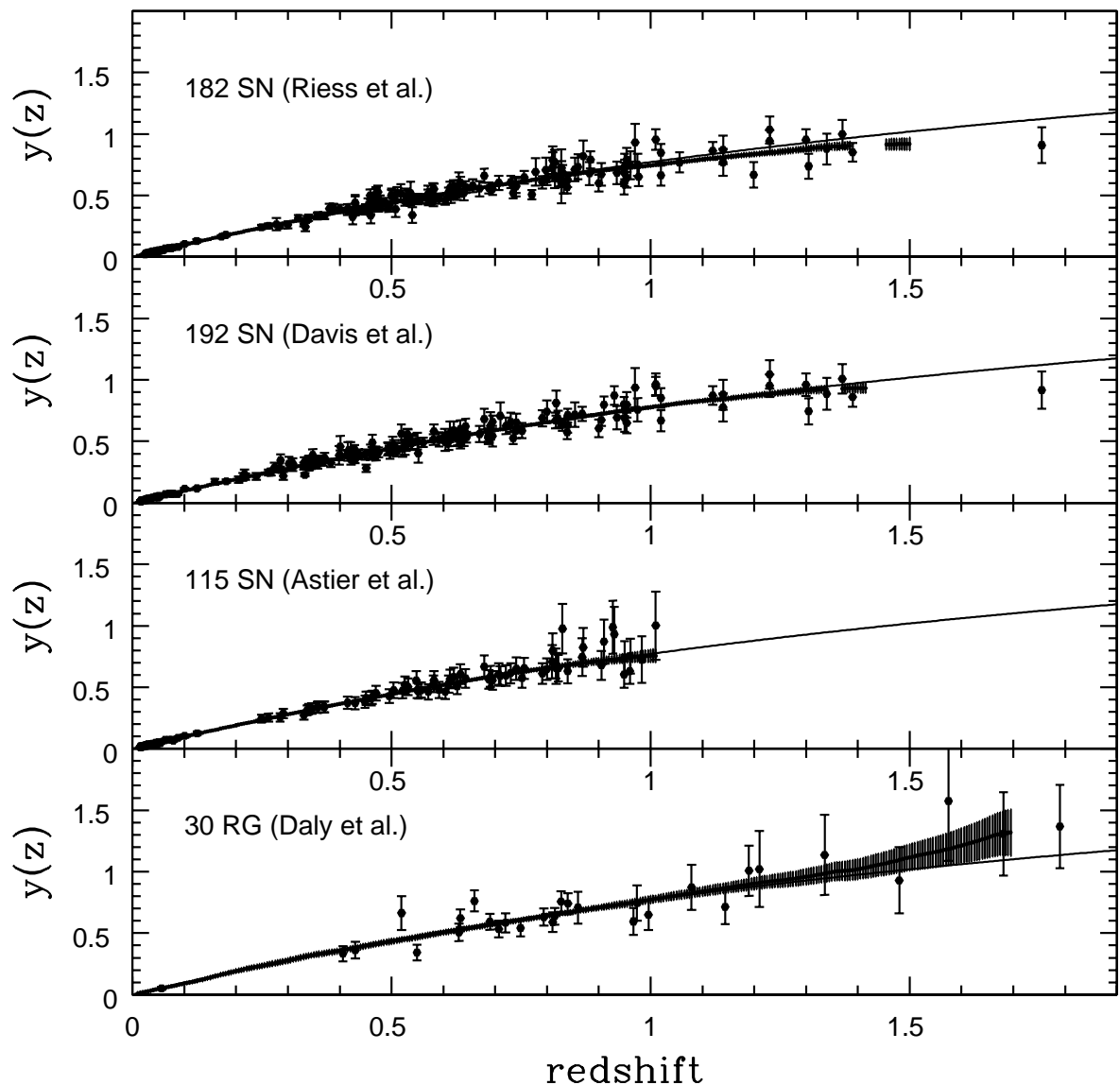}
\caption{The dimensionless coordinate distances to each source  
show along with the best fit y(z) and its one sigma error bar for
the 182 gold supernovae from Riess et al. (2007), the 192 supernovae from 
Davis et al. (2007), the 115 supernovae of Astier et al. (2006), and the 
30 radio galaxies from Daly et al. (2007).
In this and in all subsequent figures, 
the solid curve illustrates the predicted value in a standard 
LCDM model with $\Omega_m = 0.3$ and $\Omega_{\Lambda}=0.7$. } 
\label{yofz}
\end{figure}
\clearpage

\begin{figure}
\plotone{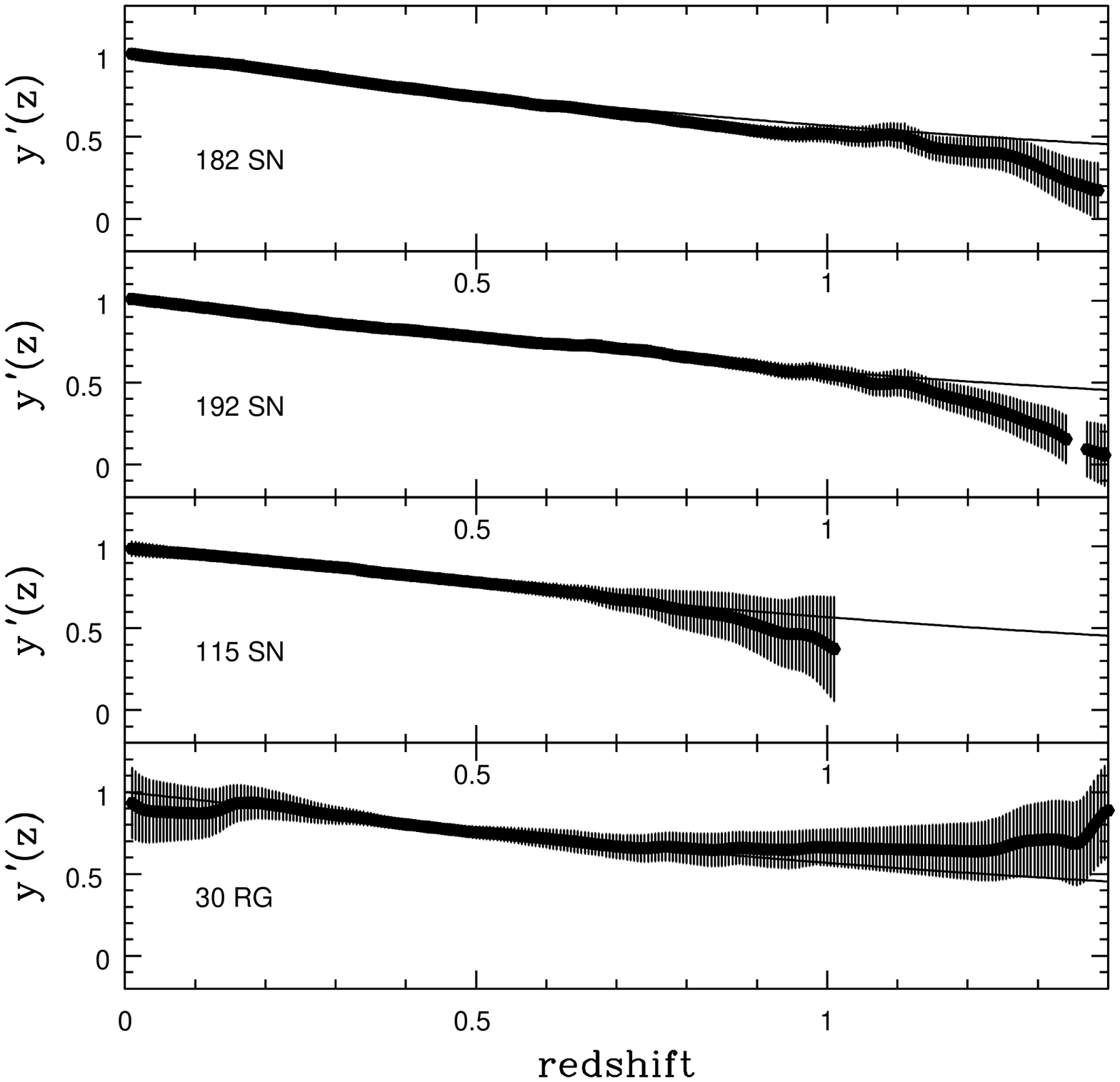}
\caption{The first derivative of the coordinate distance with respect
to redshift 
as a function of redshift for the samples described in Fig. \ref{yofz}.} 
\label{dydz}
\end{figure}
\clearpage

\begin{figure}
\plotone{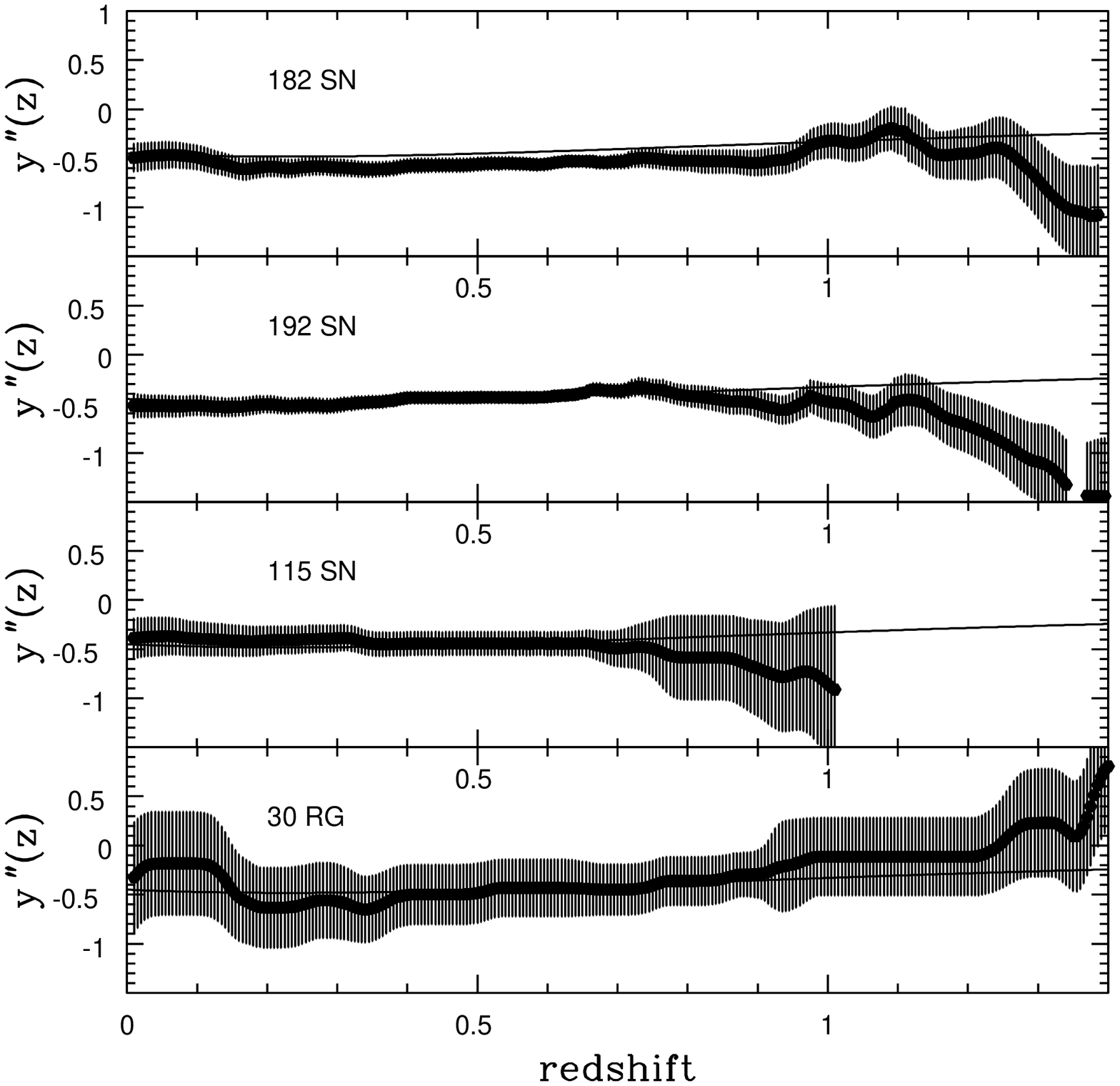}
\caption{The second derivative of the coordinate distance with respect
to redshift 
as a function of redshift for the samples described in Fig. \ref{yofz}.}  
\label{d2ydz2}
\end{figure}
\clearpage

\begin{figure}
\plotone{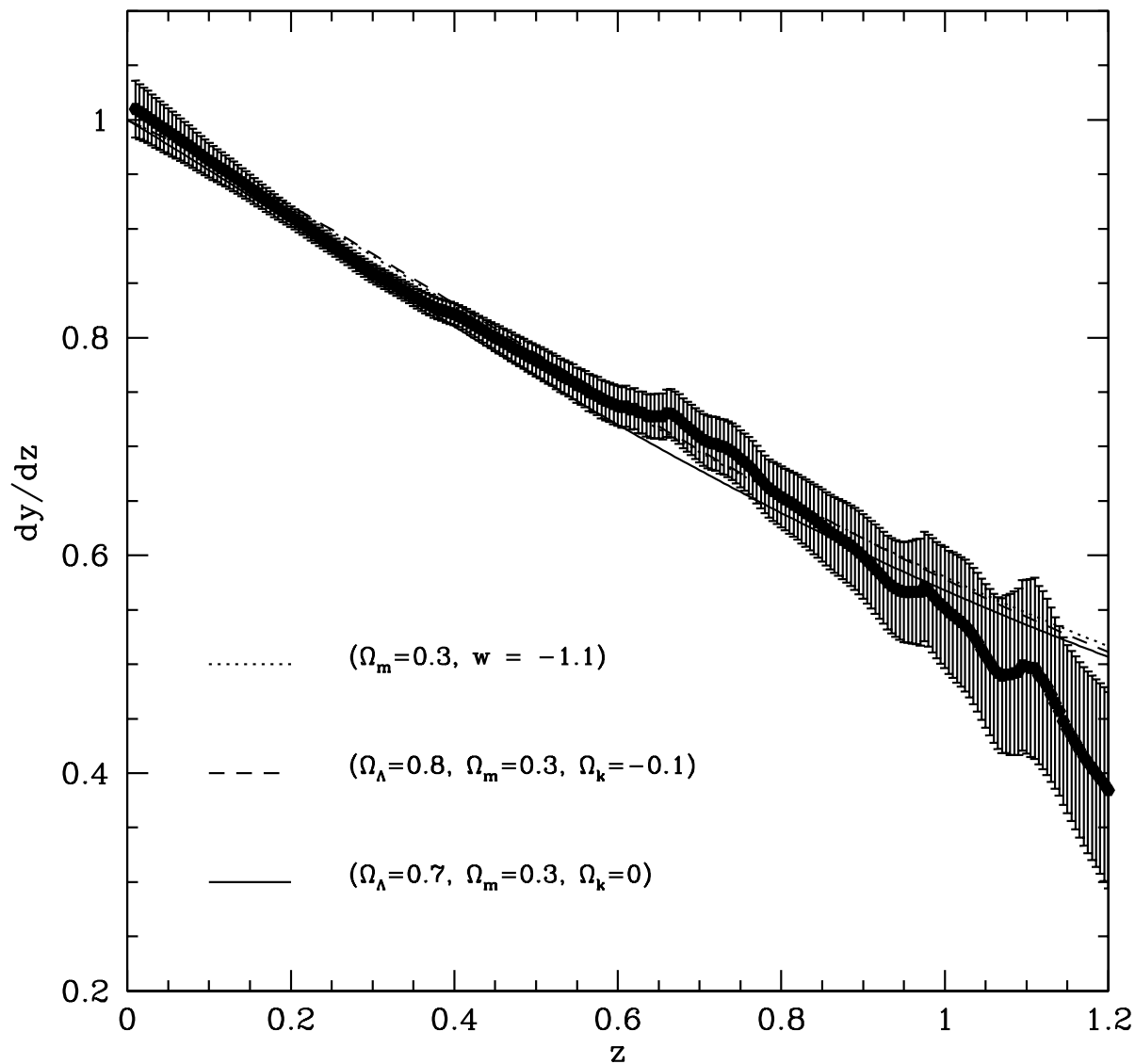}
\caption{The first derivative of the coordinate distance with 
respect to redshift as a function of 
redshift for the Davis et al. (2007) sample of 192 supernovae.  
Lines illustrating the predicted values of $y^{\prime} (z)$
for the best fit parameters obtained fitting to 
a spatially flat quintessence model and 
a lambda model with space
curvature, as well as a standard flat LCDM model, are shown. } 
\label{dydz192}
\end{figure}
\clearpage

\begin{figure}
\plotone{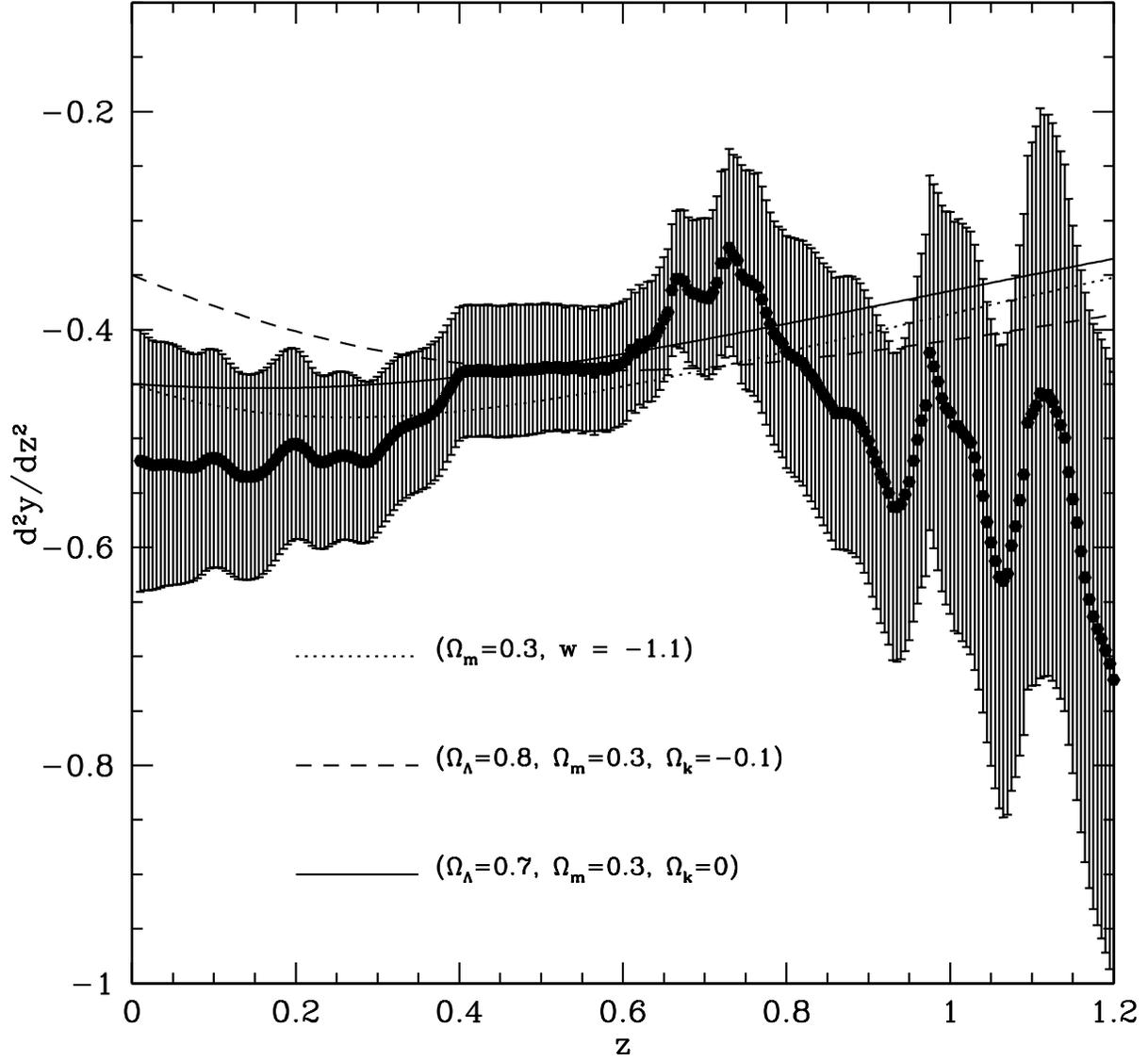}
\caption{As in Figure \ref{dydz192} for the second derivative of the 
coordinate distance with respect to redshift. } 
\label{d2ydz2192}
\end{figure}
\clearpage

\begin{figure}
\plotone{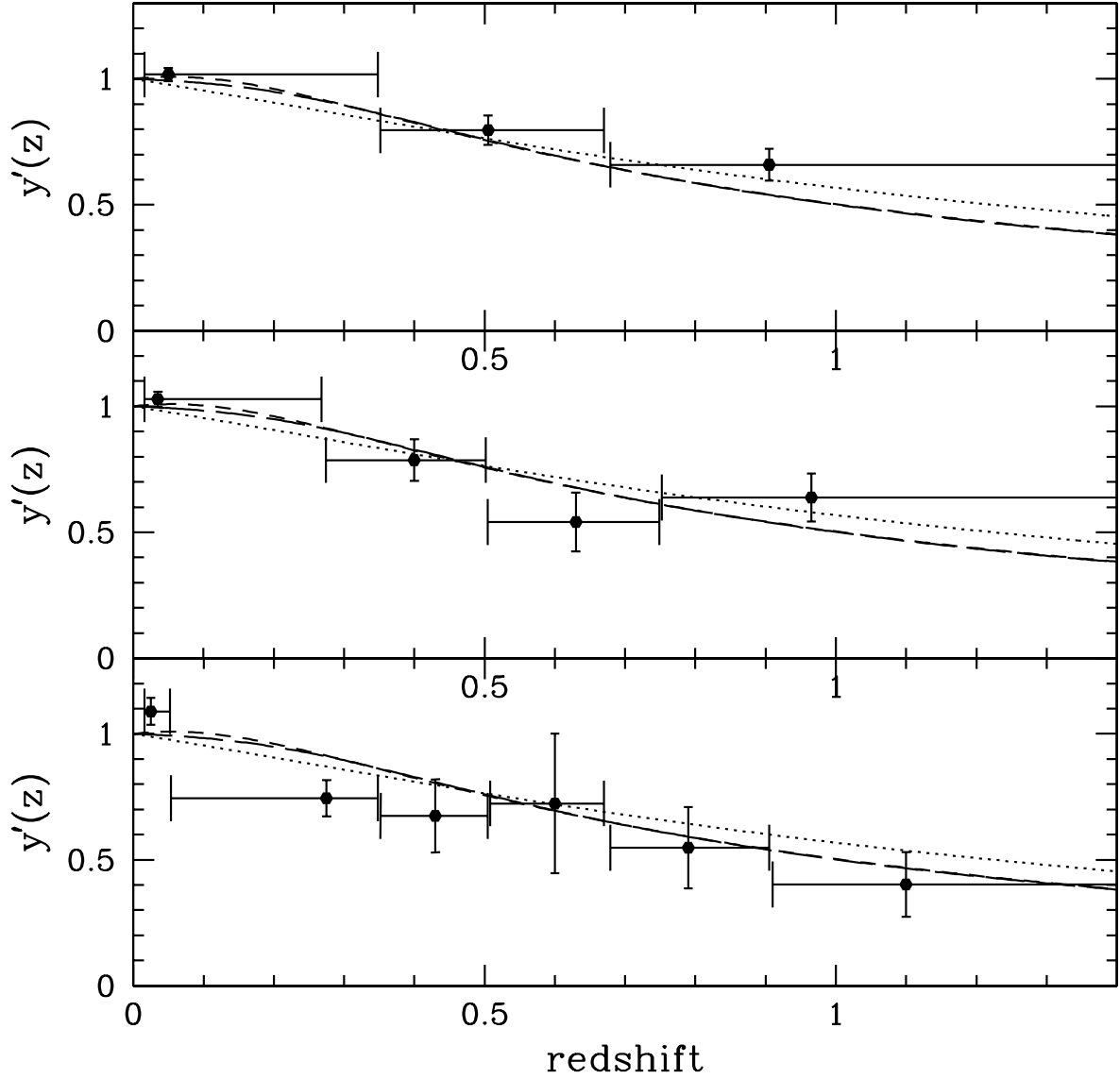}
\caption{Results obtained for the first derivative of the coordinate distance
with respect to redshift for the combined sample of 192 supernovae
and 30 radio galaxies using data split into three bins (top panel), 
four bins (middle panel), and six bins (bottom panel). The data
point at the median redshift is shown, and the horizontal
bars indicate the redshift range of the data points in the bin. 
The standard LCDM prediction for $\Omega_m = 0.3$ is indicated
by the dotted line, and the curves predicted by the 
Cardassian model and the 
generalized Chaplygin gas model, which yield nearly identical 
results, are shown by the short and long dashed curves. }
\label{yprimebin}
\end{figure}
\clearpage

\begin{figure}
\plotone{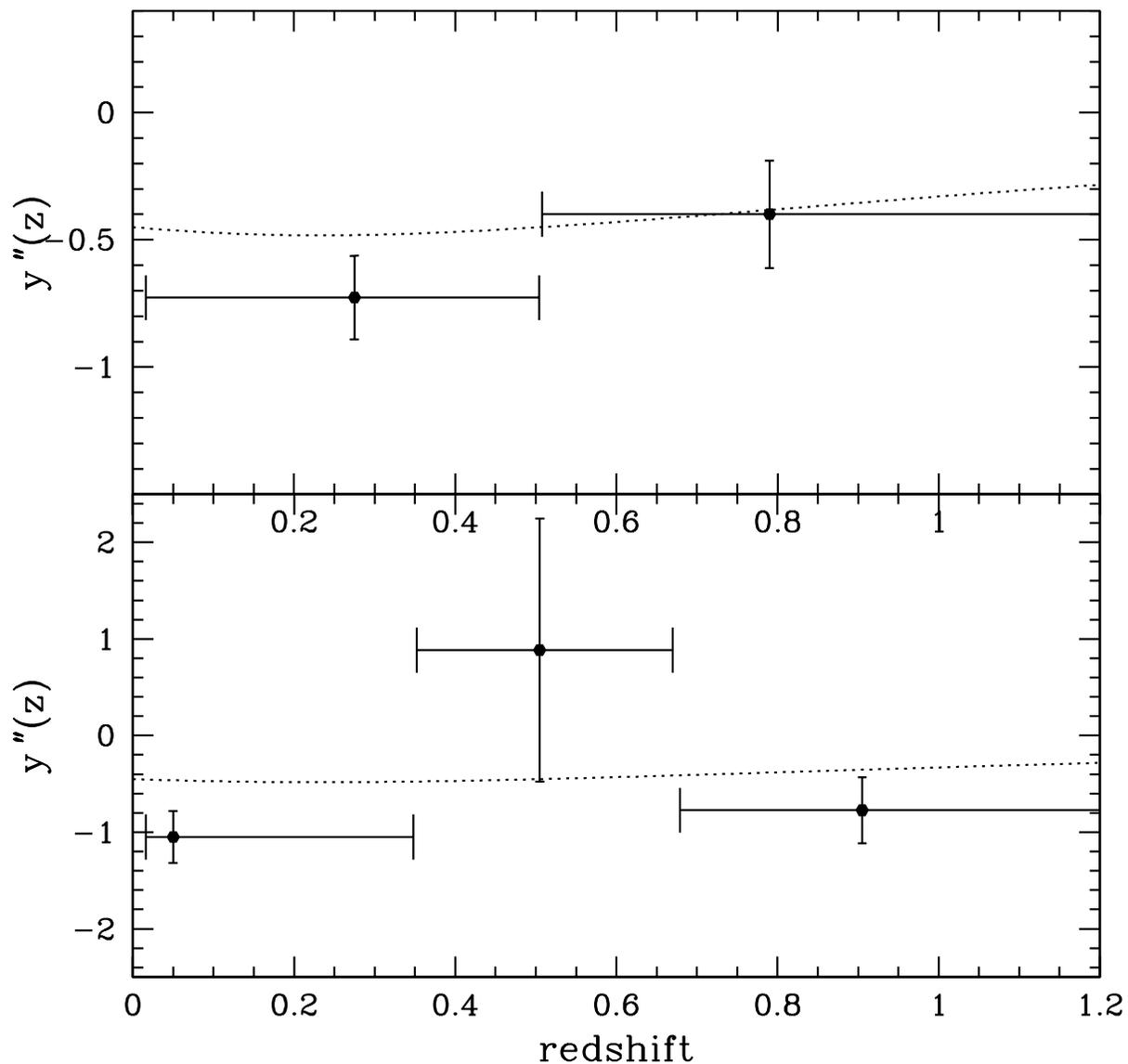}
\caption{Results obtained 
for the second derivative of the coordinate distance
with respect to redshift for the combined sample of 192 supernovae
and 30 radio galaxies using data split into two bins (top panel) 
and three (bottom panel). The data
point at the median redshift is shown, and the horizontal
bars indicate the redshift range of the data points in the bin.
The standard LCDM prediction for $\Omega_m = 0.3$ is indicated
by the dotted line. }
\label{yprimeprimebin}
\end{figure}
\clearpage

\begin{figure}
\plotone{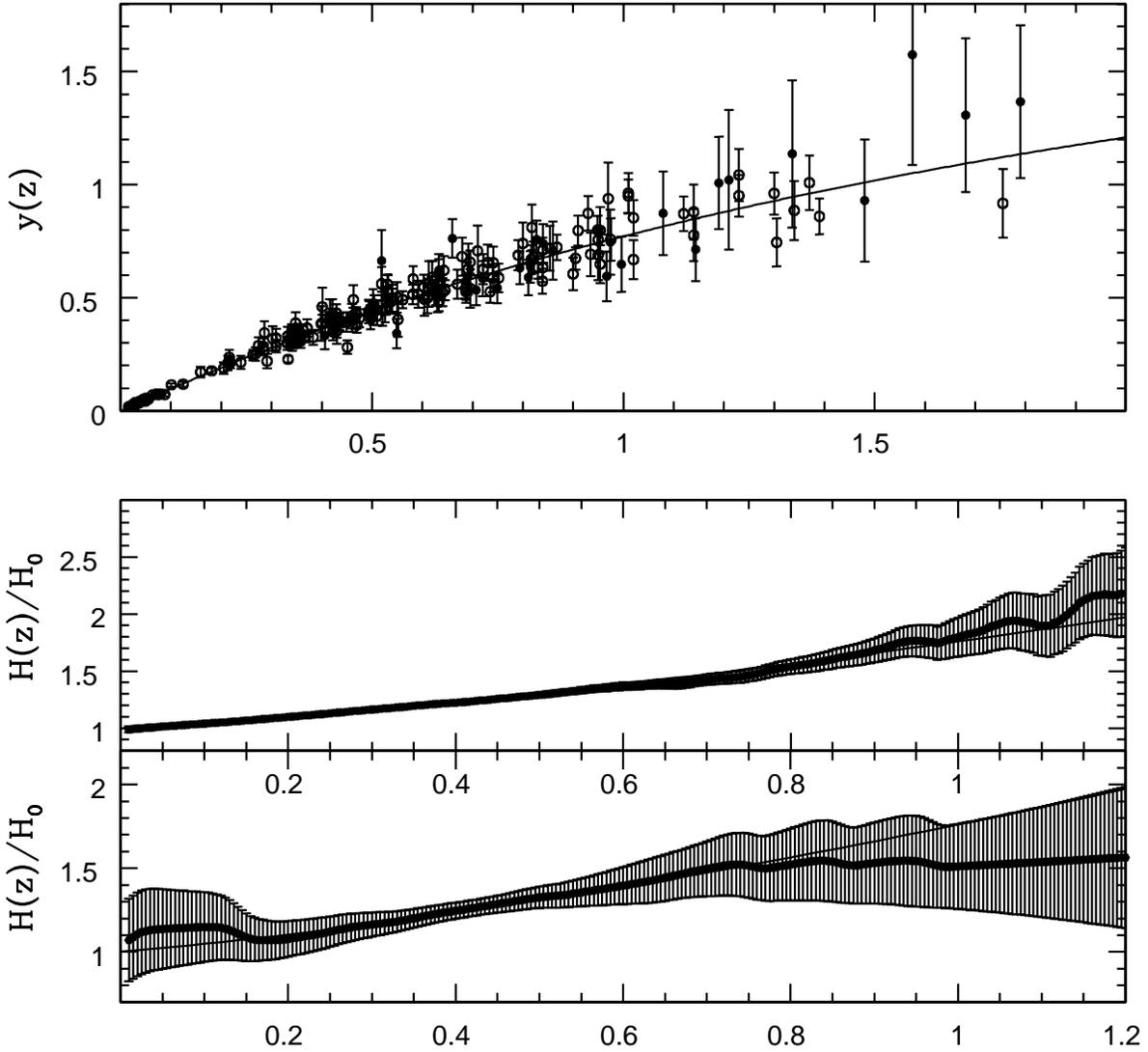}
\caption{The dimensionless coordinate distances to the 192
supernovae of Davis et al. (2007) (open circles) and the 
30 radio galaxies of Daly et al. (2007) (closed circles) are
shown in the top panel.  Our model-independent 
determination of H(z), obtained assuming only a FRW metric and
zero space curvature, is shown for the combined sample of
192 supernovae and 30 radio galaxies in the second panel,
while that for the 30 radio galaxies alone is shown in the bottom
panel. } 
\label{yEsnprg}
\end{figure}
\clearpage

\begin{figure}
\plotone{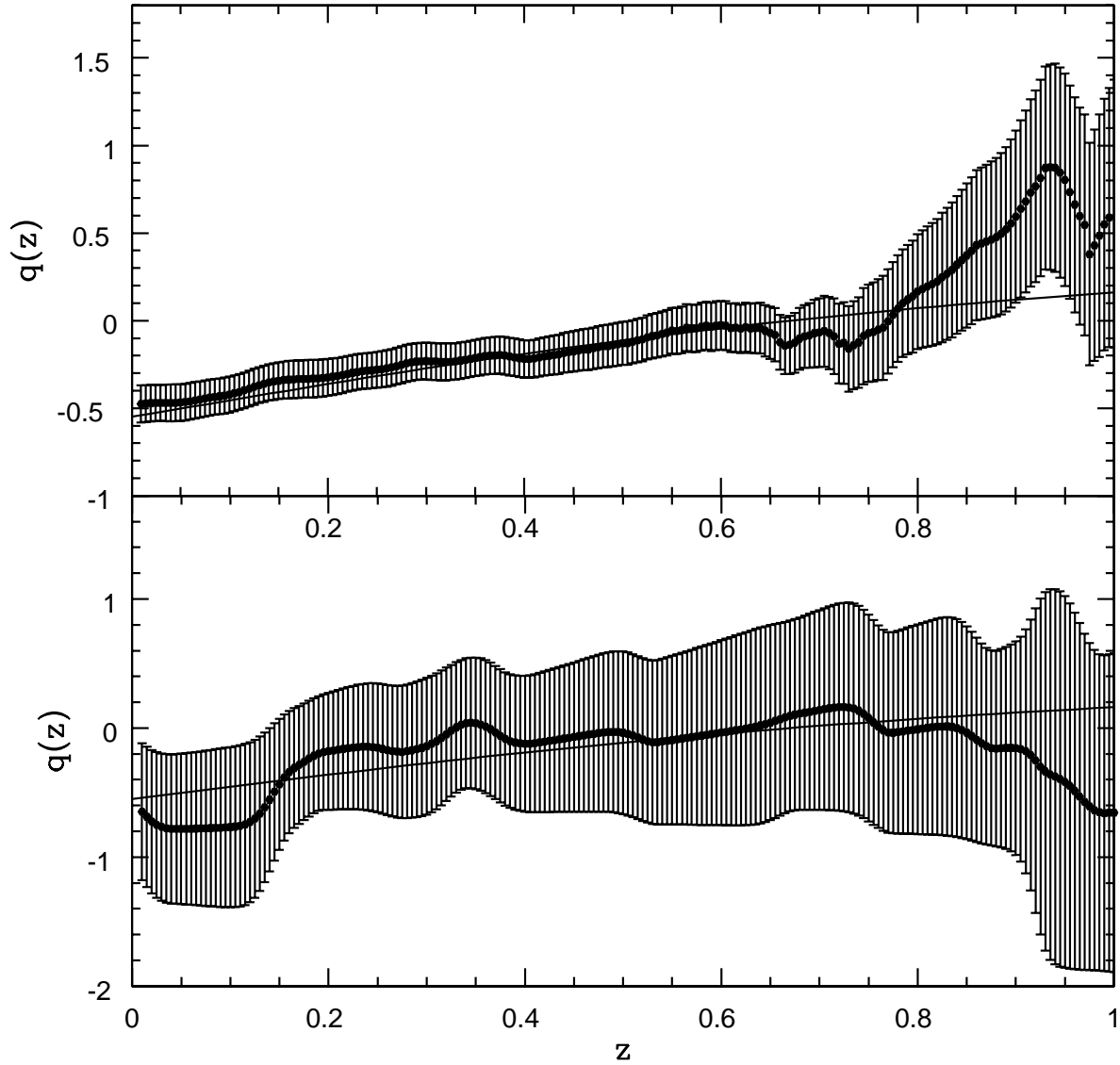}
\caption{Our model-independent 
determination of q(z), obtained assuming only a FRW metric and
zero space curvature, is shown for the combined sample of
192 supernovae and 30 radio galaxies in the top panel and
for the 30 radio galaxies alone in the bottom
panel. } 
\label{qsnprg}
\end{figure}
\clearpage

\begin{figure}
\plotone{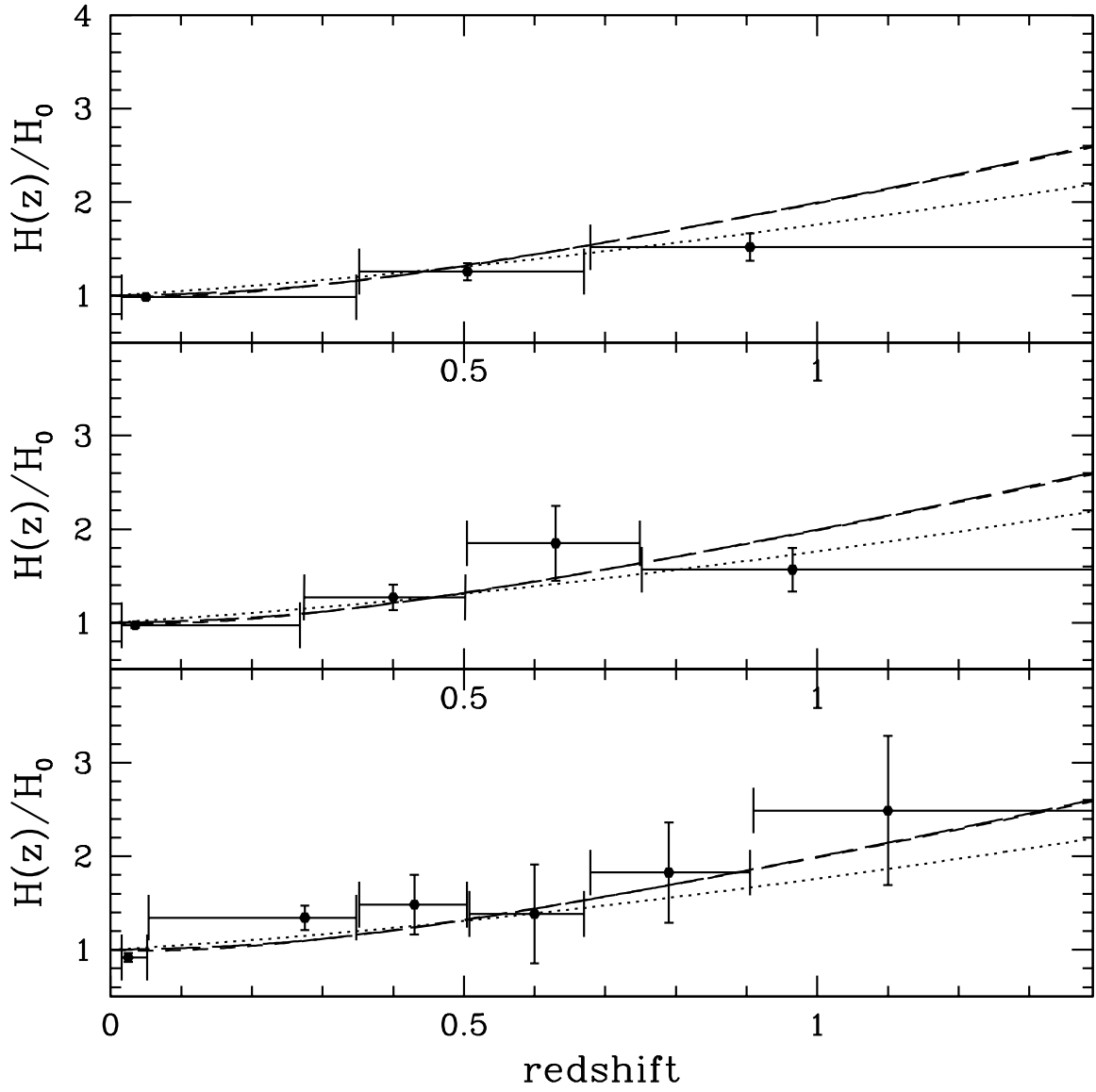}
\caption{As in Figure \ref{yprimebin} but for $H(z_{med})/H_0$.}
\label{binH}
\end{figure}
\clearpage

\begin{figure}
\plotone{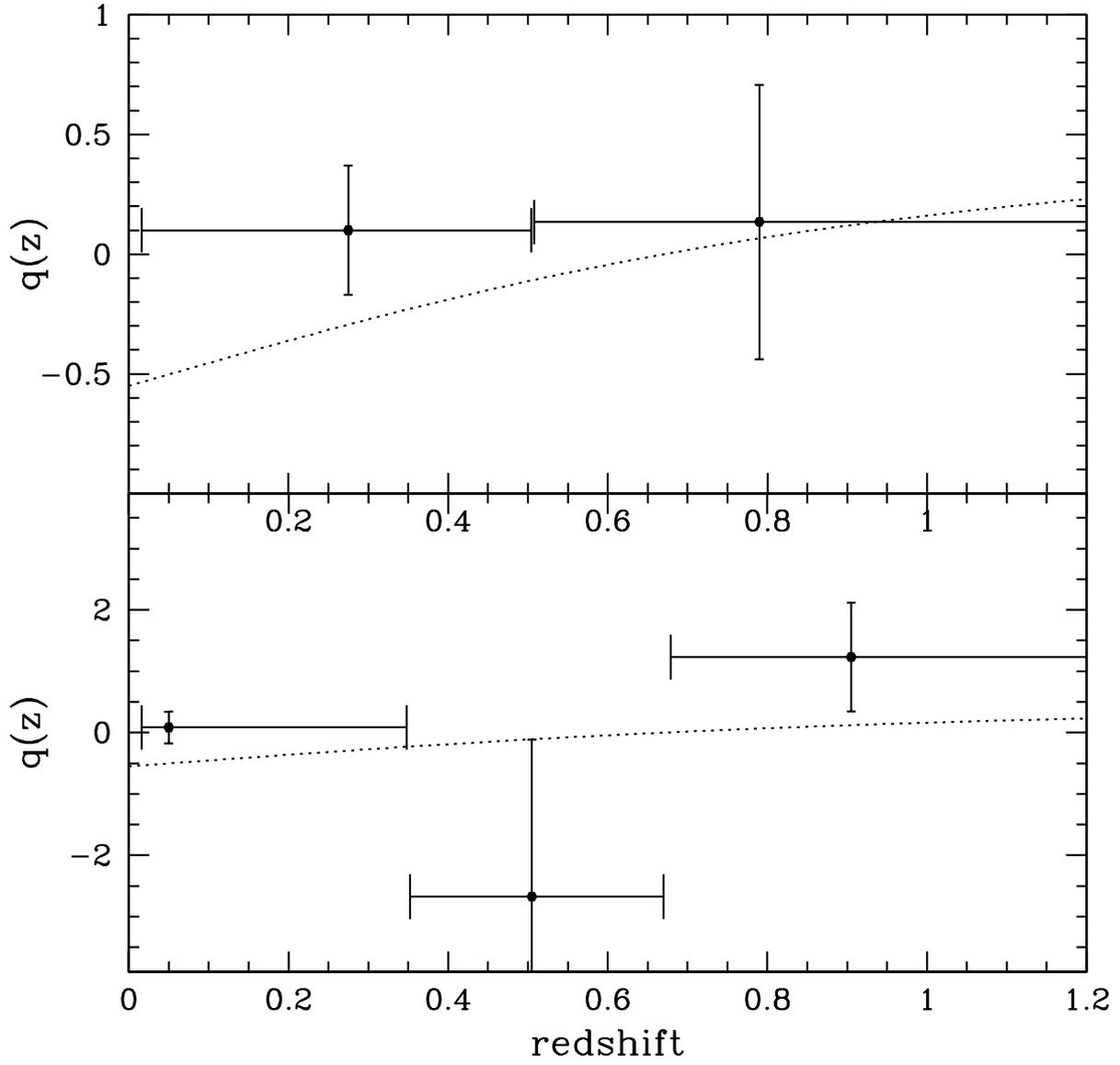}
\caption{As in Figure \ref{yprimeprimebin} but for $q(z_{med})$.}
\label{binq}
\end{figure}
\clearpage

\begin{figure}
\plotone{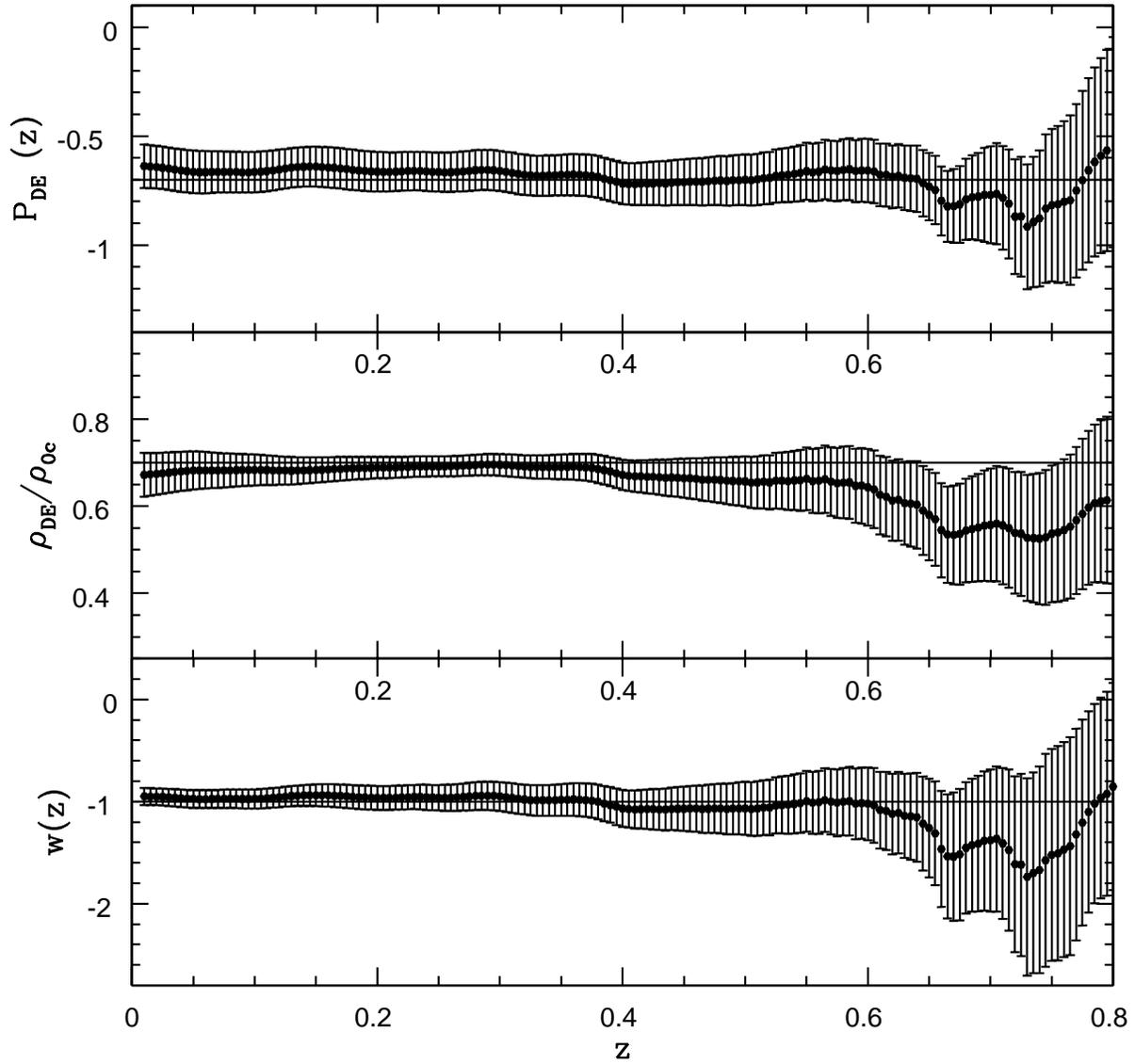}
\caption{The pressure, energy density, and equation of state of the dark
energy obtained for the combined sample of 30 radio galaxies and 
192 supernovae, 
obtained from $y^{\prime}$ and $y^{\prime \prime}$ 
assumig zero space curvature.}
\label{Pfw}
\end{figure}
\clearpage

\begin{figure}
\plotone{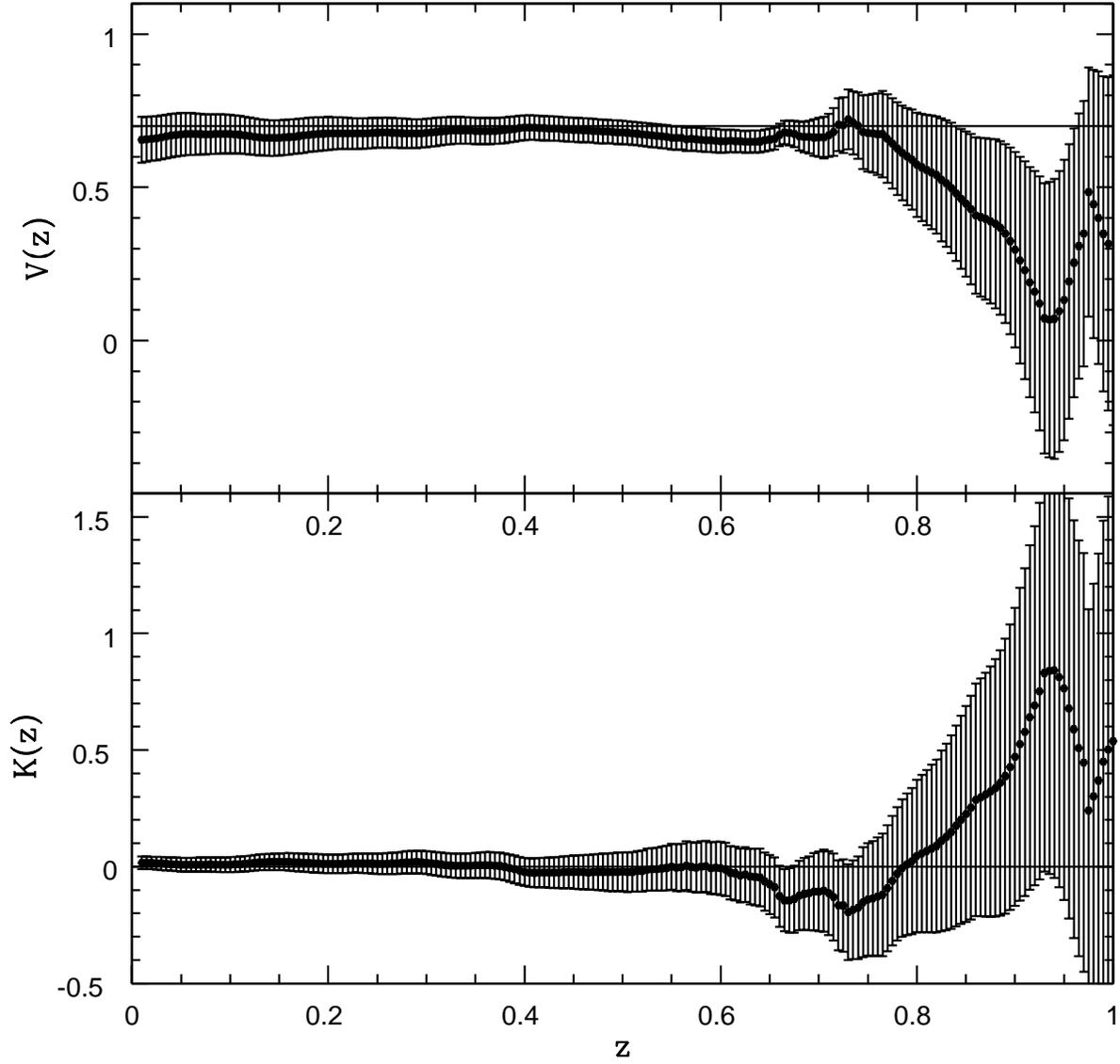}
\caption{The potential and kinetic energy density of a dark energy 
scalar field as a function of redshift for the combined sample of 
30 radio galaxies and 192 supernovae,
obtained from $y^{\prime}$ and $y^{\prime \prime}$ 
assuming zero space curvature.} 
\label{KV}
\end{figure}
\clearpage

\begin{figure}
\plotone{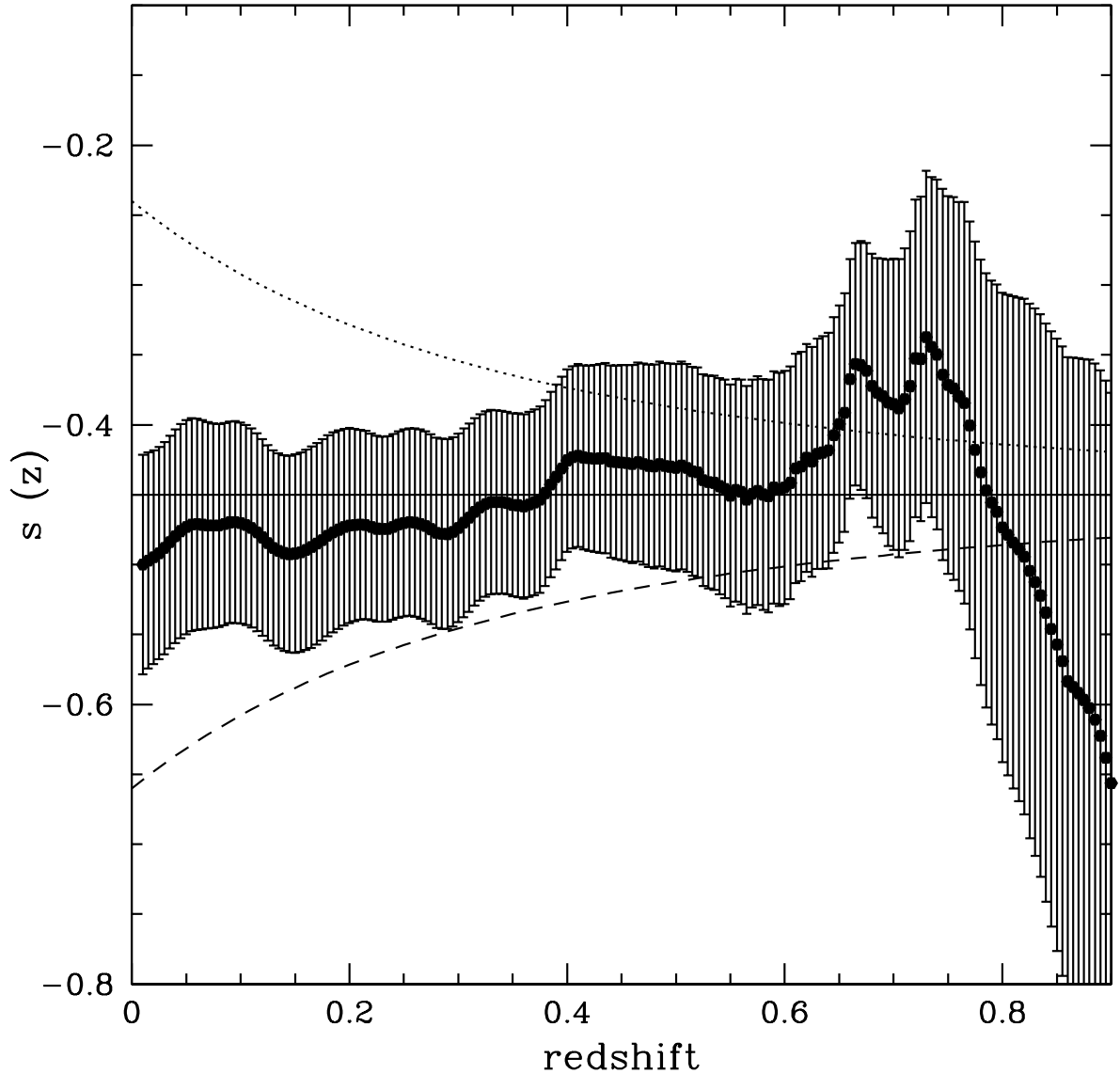}
\caption{The  
dark energy indicator for the combined sample of 30 radio 
galaxies and 192 supernovae obtained from using 
equation \ref{s}.  The behavior of $s$ predicted using
equation \ref{sp} is shown for three simple models
each assuming $\Omega_m = 0.3$, $\Omega_{DE}= 0.7$ 
and $f(z) = 1$ over the redshift range shown,
and $w = -1$ (solid line), 
$w = -0.8$ and remains constant over the redshift 
range shown (long dashed
curve), and $w=-1.2$ and remains constant over the 
redshift range shown (short dashed curve).  
If $s$ remains constant, it suggests that $w=-1$,
and the value of $s$ provides a new and indepedent
measure of $\Omega_m$.   } 
\label{S}
\end{figure}
\clearpage

\begin{figure}
\plotone{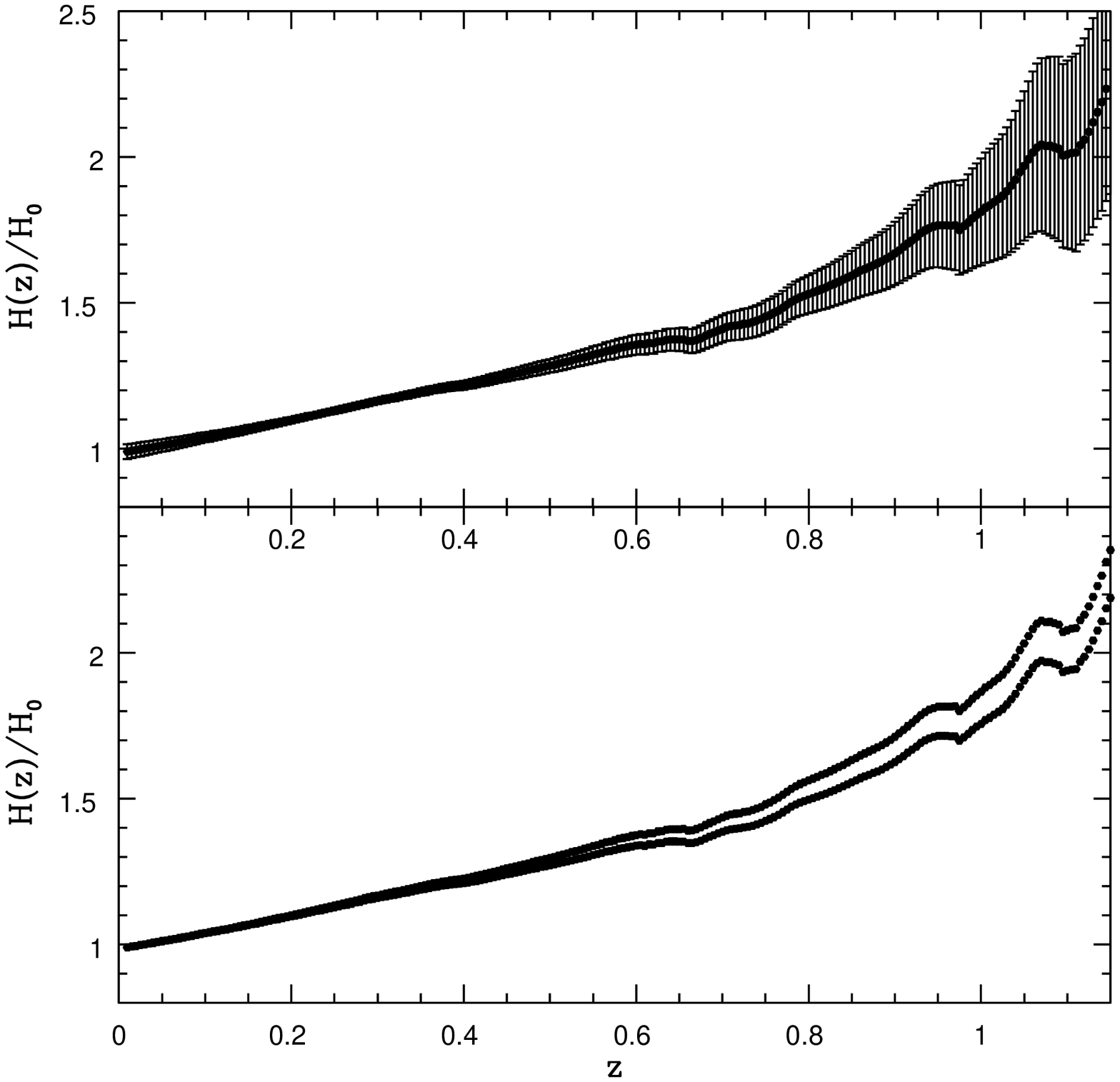}
\caption{Model independent determination of H(z) for 
the Davis et al. (2007) sample of 192 supernovae.
The top panel shows results obtained for $\Omega_k = 0$, and 
the bottom panel shows results obtained for $\Omega_k = 0.1$
(upper curve) and $\Omega_k = -0.1$ (lower curve). 
For clarity, the uncertainties are not shown on the bottom panel,
but are similar to those shown on the top panel.} 
\label{kEofz192}
\end{figure}
\clearpage

\begin{figure}
\plotone{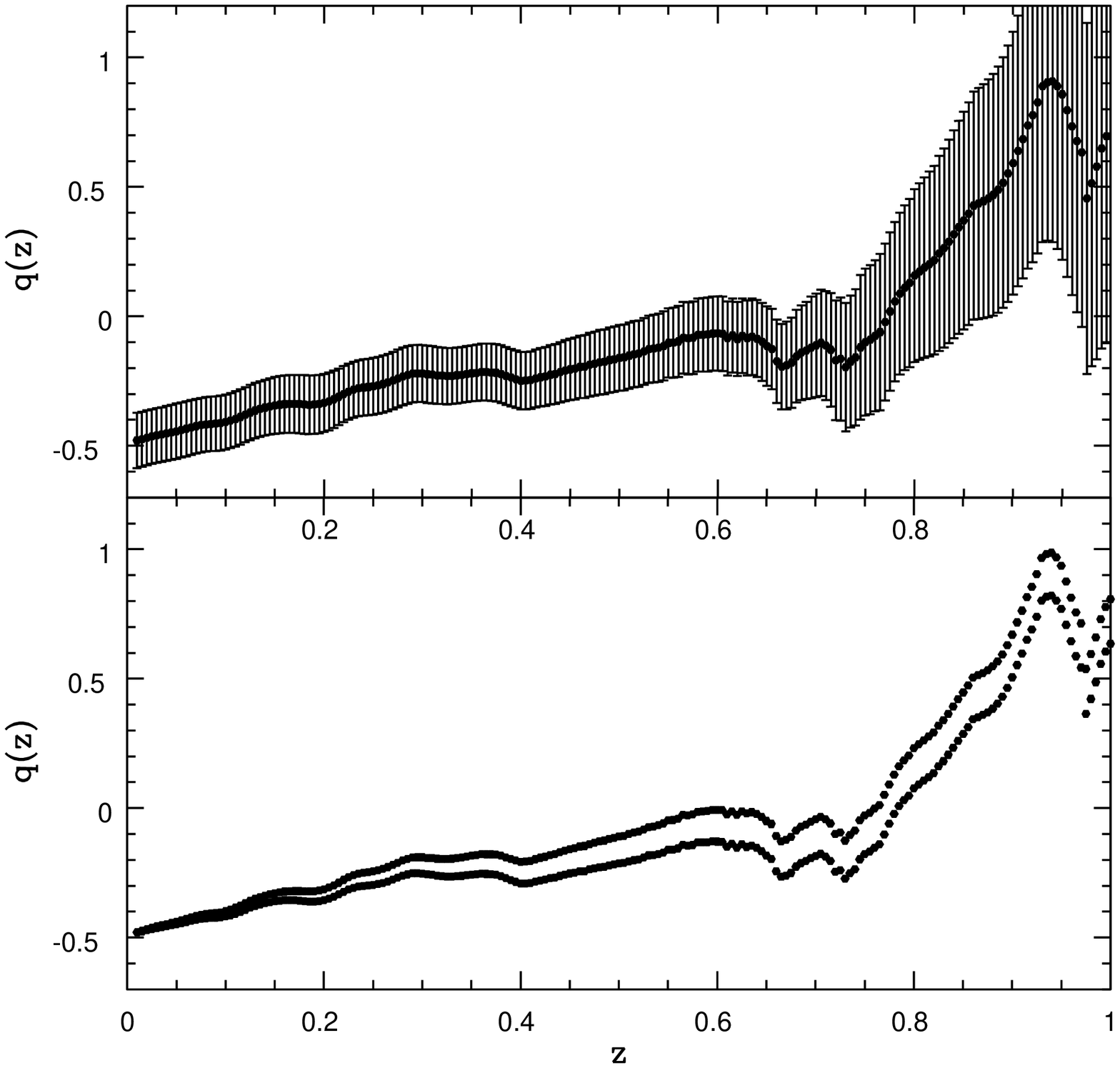}
\caption{Model independent determination of q(z) for 
the Davis et al. (2007) sample of 192 supernovae.
The top panel shows results obtained for $\Omega_k = 0$, and 
the bottom panel shows results obtained for $\Omega_k = 0.1$
(upper curve) and $\Omega_k = -0.1$ (lower curve). 
For clarity, the uncertainties are not shown on the bottom panel,
but are similar to those shown on the top panel.}
\label{kqofz192}
\end{figure}
\clearpage

\begin{figure}
\plotone{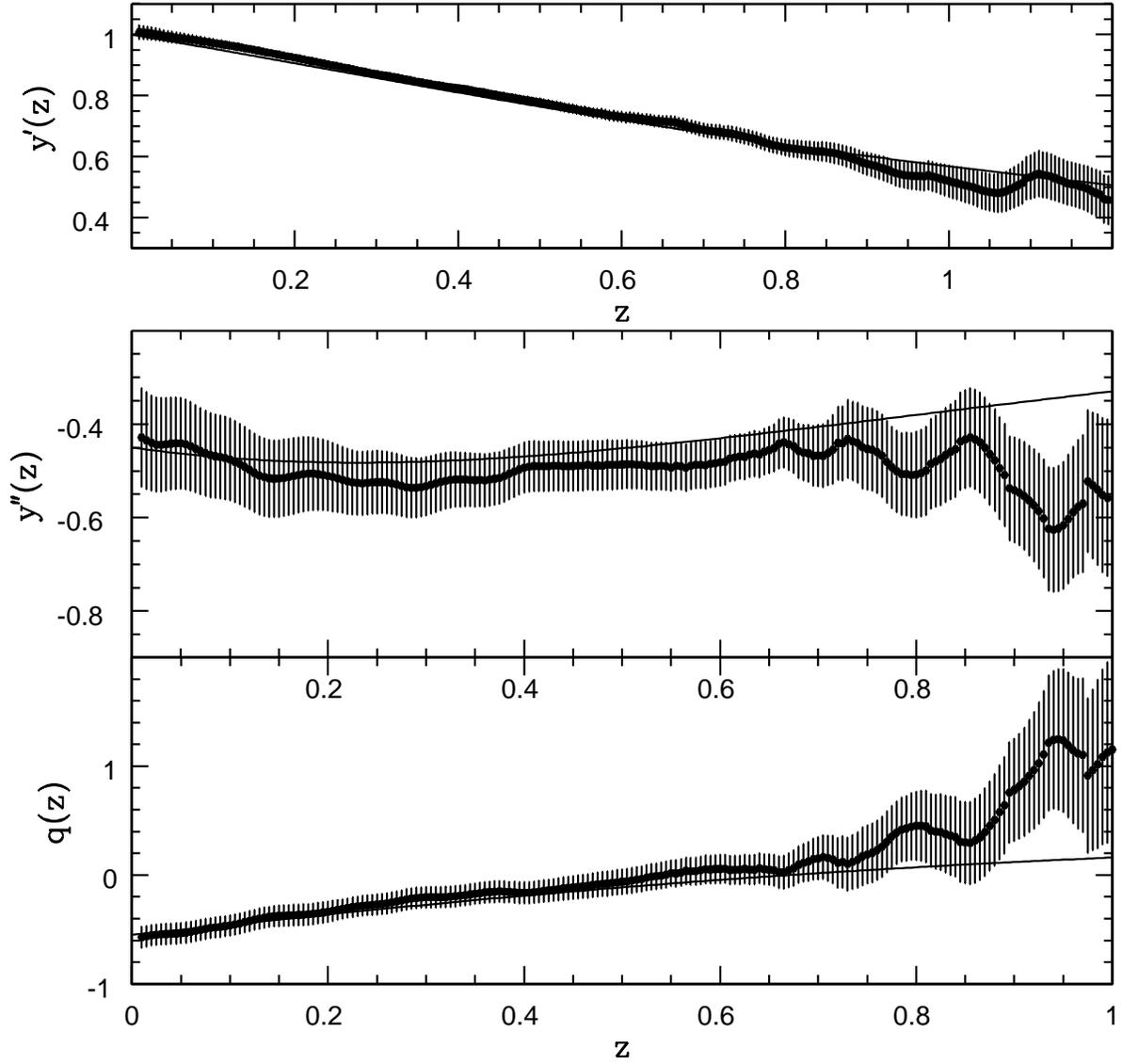}
\caption{Results for  $y^{\prime}(z)$,  
$y^{\prime \prime}(z)$, and $q(z)$ obtained with the 
combined sample of 30 radio galaxies (solid circles),
192 supernovae (open circles), and 38
SZ clusters (stars), which is shown in Fig. \ref{mudiffofz}.    
The solid curve illustrates the predicted value in a standard 
LCDM model with $\Omega_m = 0.3$ and $\Omega_{\Lambda}=0.7$.}
\label{yetal260}
\end{figure}
\clearpage

\begin{figure}
\plotone{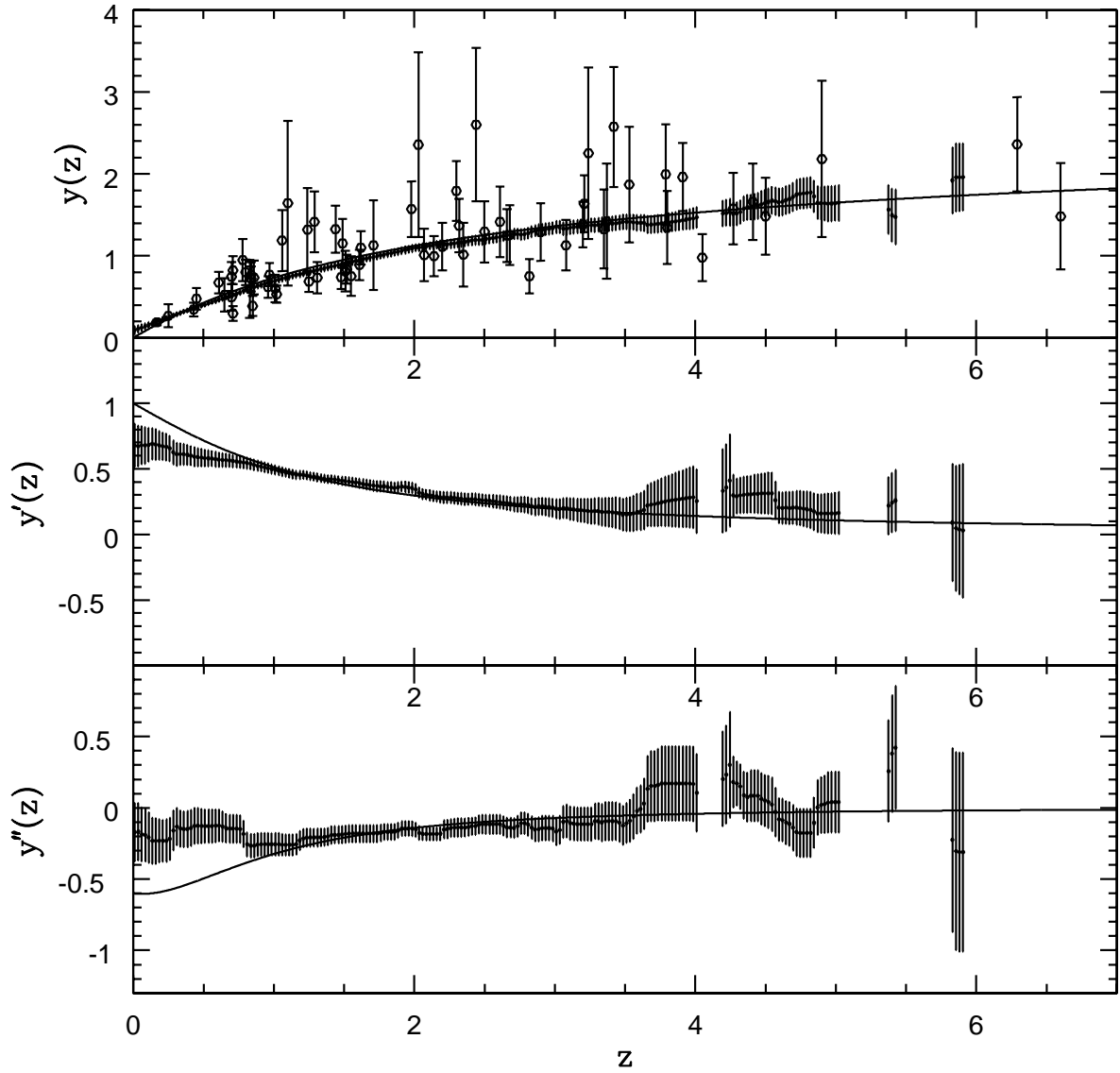}
\caption{Results for $y(z)$, $y^{\prime}(z)$, and 
$y^{\prime \prime}(z)$ obtained with the 69 gamma ray burst
data of Schaefer (2007). 
The solid curve illustrates the predicted value in a standard 
LCDM model with $\Omega_m = 0.4$ and $\Omega_{\Lambda}=0.6$,
which are the best fit values to this model obtained by Schaefer (2007). }
\label{GRB}
\end{figure}
\clearpage

\end{document}